\documentclass[11pt]{article}

\RequirePackage{etex}

\usepackage[margin=1in]{geometry}  	
\usepackage{amsfonts}
\usepackage{amsmath} 
\usepackage{booktabs}
\usepackage{multirow}
\usepackage[onehalfspacing]{setspace} 
\renewcommand{\arraystretch}{0.6} 

\usepackage{float}
\usepackage{caption}
\usepackage{graphicx}
\usepackage{adjustbox}
\usepackage{lscape}

\usepackage{adjustbox}

\usepackage{xcolor}

\usepackage{relsize}

\usepackage{bbm}
\usepackage{subfig}

\definecolor{forestgreen}{RGB}{34,139,34}
\usepackage{hyperref}
\hypersetup{
    colorlinks=true,
    linkcolor=magenta,
    filecolor=magenta, 
    citecolor=forestgreen,      
    urlcolor=blue
}

\usepackage{authblk}

\usepackage{amsthm}

\newtheorem*{proposition*}{Proposition}

\usepackage{xpatch}
\makeatletter
\xpatchcmd{\proof}{\@addpunct{.}}{\@addpunct{:}}{}{}
\makeatother

\newcommand{\overbar}[1]{\mkern 1.5mu\overline{\mkern-1.5mu#1\mkern-1.5mu}\mkern 1.5mu}

\makeatletter
\newcommand{\vast}{\bBigg@{3}}
\newcommand{\Vast}{\bBigg@{4}}
\makeatother

\makeatletter
\newcommand*{\indep}{%
  \mathbin{%
    \mathpalette{\@indep}{}%
  }%
}
\newcommand*{\nindep}{%
  \mathbin{
    \mathpalette{\@indep}{\not}
  }%
}
\newcommand*{\@indep}[2]{%
  \sbox0{$#1\perp\m@th$}
  \sbox2{$#1=$}
  \sbox4{$#1\vcenter{}$}
  \rlap{\copy0}
  \dimen@=\dimexpr\ht2-\ht4-.2pt\relax
  \kern\dimen@
  {#2}%
  \kern\dimen@
  \copy0 
} 
\makeatother

\DeclareMathOperator{\E}{\textnormal{\mbox{E}}}

\usepackage[utf8]{inputenc}
\DeclareUnicodeCharacter{200E}{}

\DeclareFontFamily{U}{mathx}{\hyphenchar\font45}
\DeclareFontShape{U}{mathx}{m}{n}{
      <5> <6> <7> <8> <9> <10>
      <10.95> <12> <14.4> <17.28> <20.74> <24.88>
      mathx10
      }{}
\DeclareSymbolFont{mathx}{U}{mathx}{m}{n}
\DeclareFontSubstitution{U}{mathx}{m}{n}
\DeclareMathAccent{\widecheck}{0}{mathx}{"71}
\DeclareMathAccent{\wideparen}{0}{mathx}{"75}

\usepackage{tikz}
\usetikzlibrary{positioning,shapes.geometric}

\usepackage{setspace}
\usepackage{lipsum}

\usepackage[american]{babel}
\usepackage{csquotes}
\usepackage[style=apa,sortcites=true,sorting=nyt,backend=biber,uniquelist=false]{biblatex}
\DeclareLanguageMapping{american}{american-apa}
\usepackage[stable]{footmisc}

\usepackage[stable]{footmisc}

\usepackage{rotating}

\makeatletter
\def\@hangfrom#1{\setbox\@tempboxa\hbox{{#1}}%
      \hangindent 0pt
      \noindent\box\@tempboxa}
\makeatother

\makeatletter
\def\@seccntformat#1{\@ifundefined{#1@cntformat}%
   {\csname the#1\endcsname\quad}  
   {\csname #1@cntformat\endcsname}
}
\let\oldappendix\appendix 
\renewcommand\appendix{%
    \oldappendix
    \newcommand{\section@cntformat}{\appendixname~\thesection\quad}
}
\makeatother

\usepackage[absolute,showboxes]{textpos}

\TPMargin{5pt}

\usepackage{thmtools, thm-restate}

\usepackage{datetime}
\def\paperversionmajor{4}
\def\paperversionminor{0}

\usepackage{sectsty}
\subsectionfont{\noindent\textbf\itshape}
\subsubsectionfont{\normalfont\itshape}

\begin{document}


\title{\textbf{Cluster randomized trials designed to support generalizable inferences} \vspace*{0.3in} }


\author[1,2]{Sarah E. Robertson}
\author[3]{Jon A. Steingrimsson}
\author[1,2,4]{Issa J. Dahabreh 
}

\affil[1]{CAUSALab, Harvard T.H. Chan School of Public Health, Boston, MA}
\affil[2]{Department of Epidemiology, Harvard T.H. Chan School of Public Health, Boston, MA}
\affil[3]{Department of Biostatistics, Brown University School of Public Health, Providence, RI }
\affil[4]{Department of Biostatistics, Harvard T.H. Chan School of Public Health, Boston, MA}

\date{}
\maketitle{}

\textbf{Running header:} Cluster trials designed for generalizability 

\textbf{Declaration:} The authors have no conflicts of interest to report. 

\newpage
\thispagestyle{empty}

\vspace*{1in}
{\Huge \centering Cluster randomized trials designed to support generalizable inferences \par }

\vspace{1in}
\noindent
\textbf{Running head:} Cluster trials designed for generalizability

\thispagestyle{empty}

\clearpage

\thispagestyle{empty}

\vspace*{1in}

\begin{abstract}
\noindent
\linespread{1.7}\selectfont
\textbf{Background:} When planning a cluster randomized trial, evaluators often have access to an enumerated cohort representing the target population of clusters. Practicalities of conducting the trial, such as the need to oversample clusters with certain characteristics to improve trial economy or to support inference about subgroups of clusters, may preclude simple random sampling from the cohort into the trial, and thus interfere with the goal of producing generalizable inferences about the target population.
\textbf{Methods:} We describe a nested trial design where the randomized clusters are embedded within a cohort of trial-eligible clusters from the target population and where clusters are selected for inclusion in the trial with known sampling probabilities that may depend on cluster characteristics (e.g., allowing clusters to be chosen to facilitate trial conduct or to examine hypotheses related to their characteristics). We develop and evaluate methods for analyzing data from this design to generalize causal inferences to the target population underlying the cohort.  \textbf{Results:} We present identification and estimation results for the expectation of the average potential outcome and for the average treatment effect, in the entire target population of clusters and in its non-randomized subset. In simulation studies we show that all the estimators have low bias but markedly different precision. \textbf{Conclusions:}  Cluster randomized trials where clusters are selected for inclusion with known sampling probabilities that depend on cluster characteristics, combined with efficient estimation methods, can precisely quantify treatment effects in the target population, while addressing objectives of trial conduct that require oversampling clusters on the basis of their characteristics.\\

\vspace{0.2in}
\noindent
\textbf{Keywords:} design, generalizability, transportability, cluster randomized trials, causal inference, interference.

\end{abstract}

\clearpage
\section{Introduction}
\setcounter{page}{1}

When conducting cluster randomized trials evaluators often have access to an enumerated cohort representing the target population of clusters. For example, in the applications motivating our work -- cluster randomized trials of vaccine effectiveness in U.S. nursing homes \parencite{crt_trial1,crt_trial2,crt_trial3} -- evaluators can often identify a roster of trial-eligible nursing homes using routinely collected data. Similarly, when conducting educational experiments, administrative data can be used to compile a list of all trial-eligible schools in a state \parencite{tipton2013stratified}. When clusters can be selected for participation from an enumerated cohort, the inferential goals of the trial and practicalities related to research economy may conflict with the goal of producing generalizable inferences for the target population underlying the enumerated cohort. For example, evaluators may be interested in oversampling certain groups of clusters to increase the trial's ability to test hypotheses about heterogeneity of treatment effects (i.e., effect modification or moderation) or to ensure that the trial can produce reasonably precise estimates in cluster subgroups defined by baseline characteristics. Furthermore, some clusters may be oversampled if they have attributes that facilitate the conduct of the trial (e.g., have infrastructure that facilitates data collection), particularly when resources are constrained. In such cases, it is possible to select clusters for participation in the trial using sampling probabilities that depend on baseline characteristics and are under the control of evaluators -- and thus, known by design. Such sampling can help achieve the goals of the trial while supporting the generalizability of inferences to the target population. 

Most recent work on generalizability methods has focused on individually randomized trials where participants are not representative of the target population, and where evaluators do not have control over an individual's decision to participate in the trial and access to an enumerated cohort of individuals is not common \parencite{cole2010, westreich2017,rudolph2017,  dahabreh2018generalizing, dahabreh2020transportingStatMed}. Some work in educational and welfare policy research has discussed generalizability analyses with cluster randomized trial data in settings where cluster participation in the trial was \emph{not} under evaluator control (e.g., \textcite{tipton2013improving, omuirch2014generalizing, tipton2017implications}). Recent work in educational research has mainly considered trials in which clusters are selected for participation by sampling within strata defined by effect modifiers, such that the average treatment effect in the trial may directly generalize to the target population \parencite{tipton2013stratified, tipton2014sample, tipton2017design}. This work has mentioned, without providing details, the possibility of using simple weighting \parencite{stuart2011use} or stratification \parencite{tipton2013improving} estimators to adjust for imbalances between the sampled clusters and the target population (e.g., for effect modifiers that were not stratified on) or when clusters are sampled with unequal probabilities across strata (e.g., to optimally estimate stratum-specific treatment effects \parencite{tipton2019designing}). 

Prior generalizability work using cluster randomized trial data, whether participation of the clusters in the trial was under the control of evaluators or not, aggregated individual-level information to the cluster level and used weighting or stratification methods to generalize inferences to the target population by estimating average treatment effects, in the target population of clusters \parencite{tipton2013improving, omuirch2014generalizing, tipton2017implications, tipton2013stratified, tipton2014sample,tipton2017design, tipton2019designing,tipton2018review}. These approaches may be inefficient because they ignore information on the relationship of covariates and treatment with the outcome, available from randomized clusters at the end of the trial (but often unavailable from non-randomized clusters). Instead, these approaches use cluster-level baseline covariate information from only non-randomized clusters via the sampling probability (or the probability of trial participation when not under the evaluators' control). By modeling the relationship of covariates and treatment with the outcome using either individual-level or cluster-level data, efficiency can be improved using an augmented inverse probability weighting estimator, without introducing bias even if the outcome model is misspecified, as long as the sampling probability is correctly specified (when the sampling probability is under investigator control, a model for it can always be correctly specified). Of note, generalizability analyses that use individual-level data require accounting for various forms of within-cluster dependence \parencite{balzer2019new}, including causal interference (e.g., due to herd immunity effects \parencite{halloran1995causal}), even if the clusters are assumed to be independent -- an issue that is well-appreciated for analyses of cluster randomized trials \parencite{murray1998design, donner2000design}.

Here, we combine recent advances in the analysis of cluster randomized trials \parencite{balzer2019new, balzer2022defining} and prior work on generalizability analyses for individually randomized trials  \parencite{dahabreh2018generalizing, dahabreh2020transportingStatMed} to propose augmented weighting estimators for analyzing cluster randomized trials where the evaluators have sampled participating clusters using known sampling probabilities that depend on baseline characteristics. The methods we describe can be used to estimate causal effects in the target population from which participating clusters are sampled, while exploiting individual-level information on the relationship of covariates and treatment with the outcome and allowing for arbitrary within-cluster dependence. We show that knowledge of the sampling probabilities leads to augmented weighting estimators that are robust to misspecification of the outcome model. In addition to robustness, we show that knowledge of the sampling probabilities can be used to develop more efficient estimators of causal estimands that pertain to the non-randomized subset of the target population, but not those that pertain to the entire target population. We evaluate the finite-sample performance of the methods in a simulation study motivated by a cluster randomized trial of vaccine effectiveness in U.S. nursing homes. Our simulations show that the precision of non-augmented weighting estimators can be improved substantially by estimating the sampling and treatment probabilities (even when they are known), rather than using the known probabilities. In contrast, the precision of the augmented weighting estimators is fairly similar whether using the true or estimated probabilities, and superior to non-augmented weighting estimators when the models for estimating the probabilities include covariates in addition to those used in the design (e.g., as may be necessary in trials with modest sample sizes that have baseline imbalances or are not representative of the target population with respect to variables not used in the design).

\section{Study design, data, and causal quantities of interest}\label{sec:data_causal_quant}

\subsection*{Study design and data} Consider the cluster version of the nested trial design for analyses extending inferences from a individually randomized trial to a target population \parencite{dahabreh2021study}: among a cohort of clusters sampled from the target population (e.g., a cohort of trial-eligible clusters), a subset is chosen to participate in a randomized trial with cluster-level treatment assignment. We assume that participation of clusters in the trial is fully under the control of the investigators (e.g., as might be the case for interventions with favorable risk-benefit profiles when all clusters in the enumerated cohort can be potentially included in the trial). 

We index clusters in the cohort of trial-eligible clusters by $j \in \{1, \ldots, m\}$; the $j$th cluster has sample size $N_j$, and we allow the sample size to vary across clusters. Individuals in cluster $j$ are indexed by $i \in \{1, \ldots, N_j\}$. We use $S_j$ for the cluster-level indicator of selection into the trial; $S_j = 1$ for randomized clusters and $S_j = 0$ for non-randomized clusters. For all clusters, both randomized and non-randomized, we have data on cluster-level covariates, $X_j$ (i.e., covariates that are constant for all individuals within a given cluster), and a matrix of individual-level covariates, $\boldsymbol{W}_j$, for all individuals in the cluster. For example, if $p$ baseline covariates are collected from each individual in cluster $j$, then $\boldsymbol{W}_j$ has dimension $N_j \times p$. The (row) components of $\boldsymbol{W}_j$ are vectors $W_{j,i}$, with dimension $1 \times p$, that contain information on individual-level covariates. Of note, various covariate ``aggregates'' across individuals in the same cluster can be included as cluster-level covariates (i.e., as elements of $X_j$), including \emph{cluster-level averages} of individual-level covariates. For example, the proportion of residents in a nursing home with a certain comorbidity can be calculated using individual-level information and the resulting derived variable can be treated as a cluster-level covariate. Throughout, we assume that the covariates $(X_j, \boldsymbol{W}_j)$ are measured at baseline, so that they cannot be affected by trial participation or treatment. 

Selection into the trial depends on sampling probabilities that are chosen by the evaluators and are allowed to depend on covariates $(X_j, \boldsymbol{W}_j)$, using a Bernoulli-type sampling scheme at the cluster level \parencite{breslow2007weighted, saegusa2013weighted, dahabreh2019generalizing}. We use $A_j$ to denote the cluster-level treatment assignment; we only consider finite sets of possible treatments, which we denote as $\mathcal A$. We use $\boldsymbol{Y}_j$ to denote the vector of individual-level outcomes, such that $\boldsymbol{Y}_j = (Y_{j,i} : i \in \{1, \ldots, N_j\})$ for each cluster $j$. We define the \emph{cluster-level average observed outcome} in cluster $j$, $\overbar{Y}_j$, as $\overbar{Y}_j = \frac{1}{N_j}\sum_{i = 1}^{N_j} Y_{j,i}.$

We assume independence across clusters, but we allow for arbitrary statistical dependence among individuals within each cluster (sometimes referred to as a partial interference assumption \parencite{hudgens2008toward}). Such dependence can occur via multiple mechanisms, including (1) \emph{shared exposures:} individuals share measured and unmeasured cluster-level factors that affect the outcome and response to treatment; (2) \emph{contagion:} occurrence of the outcome in one individual might affect the outcome of another individual in the same cluster; (3) \emph{covariate interference:} one individual's covariates may affect the outcomes of other individuals in the same cluster; or (4) \emph{treatment-outcome interference:} one individual's treatment assignment may affect the outcome of other individuals in the same cluster.

We collect data on baseline covariates from the cohort of trial-eligible clusters and view the cohort as a random sample from that target population of clusters. Treatment and outcome data are only needed from clusters participating in the trial; the observed data are independent and identically distributed realizations of the random tuple $\boldsymbol{O}_j = (X_j, \boldsymbol{W}_j, S_j, S_j \times A_j, S_j \times \boldsymbol{Y}_j)$, $j \in \{1, \ldots, m\}$.

\subsection*{Causal estimands} 

Let $Y_{j,i}^a$ be the potential (counterfactual) outcome for individual $i$ in cluster $j$ under intervention to assign treatment $a \in \mathcal A$ \parencite{rubin1974, robins2000d} and let $\boldsymbol{Y}^a_j$ denote the vector of potential outcomes in cluster $j$ under intervention to assign treatment $a \in \mathcal A$, such that $\boldsymbol{Y}^a_j = (Y_{j,i}^a : i \in \{1, \ldots, N_j\})$. We define the \emph{average potential outcome} in cluster $j$, $\overbar{Y}^{a}_j$, as $\overbar{Y}^{a}_j = \frac{1}{N_j}\sum_{i = 1}^{N_j} Y_{j,i}^a.$

Following \textcite{balzer2019new, balzer2022defining}, our causal quantity of interest is the (cluster-level) expectation of the average potential outcome in the target population of clusters, $\E\left[\overbar{Y}^{a}\right]. $ 
This expectation, over the entire target population of clusters, will be different from the expectation of the average potential outcome in the randomized subset of the target population, that is $\E\left[\overbar{Y}^{a}\right] \neq \E\left[\overbar{Y}^{a} | S = 1\right]$, when factors which affect the outcome are differentially distributed between clusters that are selected into the trial and those that are not. The average treatment effect in the target population comparing treatments $a \in \mathcal A$ and $a^\prime \in \mathcal A$ is a contrast of the corresponding expectations of the average potential outcomes $\E\left[\overbar{Y}^{a} - \overbar{Y}^{a^\prime} \right] = \E\left[\overbar{Y}^{a}\right] - \E\left[ \overbar{Y}^{a^\prime} \right]$. Also of interest are causal quantities in the non-randomized subset of the target population. Specifically, the expectation of the average potential outcome in the non-randomized subset of the target population, $\E\left[\overbar{Y}^{a} | S = 0 \right]$, and the average treatment effect comparing treatments $a \in \mathcal A$ and $a^\prime \in \mathcal A$, $\E\left[\overbar{Y}^{a} - \overbar{Y}^{a^\prime} | S = 0 \right] = \E\left[\overbar{Y}^{a} | S = 0 \right] - \E\left[\overbar{Y}^{a^\prime} | S = 0 \right]$.


\section{Identification}

\subsection*{For the target population}

\paragraph{Identifiability conditions:} The following conditions are sufficient to identify the expectation of the average potential outcome in the target population: \emph{A1. Consistency of cluster-level average potential outcomes:} if $A_j = a$, then $\overbar{Y}^{a}_j = \overbar{Y}_j$ for every $j \in \{1, \ldots, m\}$ and every $a \in \mathcal A$. \emph{A2. Conditional exchangeability over $A$ in the cluster randomized trial:} $\overbar{Y}^{a} \indep A | X, \boldsymbol{W}, S = 1$ for every $a$. \emph{A3. Positivity of treatment assignment probability in the trial:} $\Pr[A = a | X = x, \boldsymbol{W} = \boldsymbol{w}, S = 1] > 0$ for every $a$, and every $x$ and $\boldsymbol{w}$ with positive density in the trial. \emph{A4. Conditional exchangeability over $S$:} $\overbar{Y}^{a} \indep S | X, \boldsymbol{W}$ for every $a$. \emph{A5. Positivity of trial participation:} $\Pr[S = 1 | X = x, \boldsymbol{W} = \boldsymbol{w}] > 0$ for every $x$ and $\boldsymbol{w}$ with positive density in the target population. Note that conditions A2 through A5 are supported by study design when clusters are selected using sampling probabilities known to the evaluators and treatment is randomly assigned among randomized clusters. Condition A1 should be judged on the basis of substantive knowledge, but can be rendered more plausible by study design (e.g., using appropriate definitions of clusters to ensure the partial interference assumption is plausible). Of note, if evaluators are only interested in learning about average treatment effects (but not about the expectation of the average potential outcome under each treatment), identification is possible under weaker assumptions (see \parencite{dahabreh2018generalizing, dahabreh2020transportingStatMed} for analogous results in the case of individually randomized trials). We focus on identification of the expectation of the average potential outcome under each treatment because outcomes under different treatments are of inherent scientific and policy interest, and necessary for contextualizing treatment effects. 

\paragraph{Identification:} As shown in Appendix \ref{appendix:identification}, and similar to work on individually randomized trials \parencite{dahabreh2018generalizing}, under the above conditions, the expectation of the average potential outcome in the target population, $\E\left[\overbar{Y}^{a}\right]$, is identified by $$\psi(a) \equiv \E \Big[ \E\left[\overbar{Y} | X, \boldsymbol{W}, S = 1, A = a \right] \Big],$$ where the outer expectation is over the distribution of the target population of clusters. The average treatment effect in the target population can be identified by taking differences between expectations of the average potential outcomes under different treatments.

\subsection*{For the non-randomized subset of the target population}

\paragraph{Identifiability conditions:} To identify the expectation of the average potential outcome in the non-randomized subset of the target population, we retain conditions A1 through A4 and replace condition A5 by the following, slightly weaker, condition:

\vspace{0.1in}
\noindent
\emph{A5$^*$. Positivity of trial participation:} $\Pr[S = 1 | X = x, \boldsymbol{W} = \boldsymbol{w}] > 0$ for every $x$ and $\boldsymbol{w}$ with positive density among the non-randomized subset of the target population.
Condition A5$^*$ is supported by the study design when the sampling probabilities are under the control of the evaluators. 

\paragraph{Identification:} As shown in Appendix \ref{appendix:identification}, and similar to work on individually randomized trials \parencite{dahabreh2020transportingStatMed}, under identifiability conditions A1 through A4 and condition A5$^*$, the expectation of the average potential outcome in the non-randomized subset of the target population, $ \E\left[\overbar{Y}^{a} | S = 0 \right] $, is identified by $$\phi(a) \equiv \E \Big[ \E\left[\overbar{Y} | X, \boldsymbol{W}, S = 1, A = a \right] \big| S = 0 \Big].$$ The average treatment effect in the non-randomized subset of the target population can be identified by taking differences between the expectations of the average potential outcomes in the non-randomized subset of the target population under different treatments.

\section{Estimation and inference} \label{sec:estimation}

\subsection*{For the target population}

\paragraph{Estimation:} We propose the following augmented inverse probability of selection weighting estimator for $\psi(a)$:
\begin{equation}\label{eq:AIPW_est_all}
    \begin{split}
    \widehat{\psi}(a) &= \dfrac{1}{m} \sum\limits_{j = 1}^{m} \Bigg\{ \dfrac{I(S_j = 1, A_j = a)}{\widehat p(X_j, \boldsymbol{W}_j) \widehat e_a(X_j, \boldsymbol{W}_j)}  \Big\{ \overbar{Y}_j - \widehat g_a(X_j, \boldsymbol{W}_j) \Big\} + \widehat g_a(X_j, \boldsymbol{W}_j) \Bigg\},
    \end{split}
\end{equation}
where $\widehat p(X, \boldsymbol{W})$ is an estimator for $\Pr[S = 1 | X, \boldsymbol{W}] $; $\widehat e_a(X, \boldsymbol{W})$ is an estimator for $\Pr[ A  = a | X, \boldsymbol{W}, S = 1]$ (the known-by-design sampling and treatment assignment probabilities can be used instead); and $\widehat g_a(X, \boldsymbol{W})$ is an estimator for $\E\left[ \overbar{Y} | X, \boldsymbol{W}, S = 1, A = a \right]$. In Appendix \ref{appendix:influence functions}, we show that this estimator is the efficient one whether the functions $\Pr[S = 1 | X, \boldsymbol{W}] $ and $\Pr[ A = a | X, \boldsymbol{W}, S ]$ are known to the evaluators or have to be estimated. In Appendix \ref{appendix:robustness}, we show that $\widehat{\psi}(a)$ is robust in the sense that it converges to $\psi(a)$ regardless of whether the estimator $\widehat g_a(X, \boldsymbol{W})$ is consistent for $\E\left[ \overbar{Y} | X, \boldsymbol{W}, S = 1, A = a \right]$.

\paragraph{Inference:} We estimate the sampling variance of $\widehat \psi(a)$ as 
\begin{equation}
\widehat \sigma^2_{\widehat \psi(a)} = \dfrac{1}{m} \widehat{Var} \Big[ \mathit{\widehat\Psi}^1_j(a) \Big],
\end{equation}
where $ \widehat{Var} \Big[ \mathit{\widehat\Psi}^1_j(a) \Big] $ is the sample variance of the influence curve $\mathit{\widehat\Psi}^1_j(a)$ (the ``sample analog'' of the influence function \parencite{tsiatis2007} we give in Appendix \ref{appendix:influence functions}): $$ \mathit{\widehat\Psi}^1_j(a) = \dfrac{I(S_j = 1, A_j = a)}{\widehat p(X_j, \boldsymbol{W}_j) \widehat e_a(X_j, \boldsymbol{W}_j)} \Big\{ \overbar{Y}_j - \widehat g_a(X_j, \boldsymbol{W}_j) \Big\} + \widehat g_a(X_j, \boldsymbol{W}_j) - \widehat \psi(a). $$ The sampling variance can be used to obtain a $(1 - \alpha)\%$ confidence interval as $\Big( \widehat \psi(a) \pm z_{1-\alpha/2} \times \widehat \sigma_{\widehat \psi(a)} \Big),$ where $z_{1-\alpha/2}$ is the $(1-\alpha/2)$ quantile of the standard normal distribution. Alternatively, inference may also be obtained via the non-parametric bootstrap \parencite{efron1994introduction}, accounting for clustering \parencite{balzer2019new}.  

\subsection*{For the non-randomized subset of the target population}

\paragraph{Estimation:} We propose the following augmented inverse odds of selection weighting estimator for $\phi(a)$: 
\begin{equation}\label{eq:AIOW_est_S0}
    \begin{split}
    \widehat{\phi}(a) &= \left\{\sum\limits_{j = 1}^{m} I(S_j = 0)\right\}^{-1} \sum\limits_{j = 1}^{m}  \dfrac{I(S_j = 1, A_j = a) \big\{ 1- \widehat p(X_j, \boldsymbol{W}_j) \big\} }{\widehat p(X_j, \boldsymbol{W}_j) \widehat e_a(X_j, \boldsymbol{W}_j)} \Big\{ \overbar{Y}_j -  \widehat g_a(X_j, \boldsymbol{W}_j) \Big\}  \\
    &\quad\quad\quad + \left\{\sum\limits_{j = 1}^{m} I(S_j = 0)\right\}^{-1} \sum\limits_{j = 1}^{m}  \big\{ 1-\widehat p(X_j, \boldsymbol{W}_j)\big \} \widehat g_a(X_j, \boldsymbol{W}_j).
    \end{split}
\end{equation}

In Appendix \ref{appendix:influence functions}, we show that $\widehat{\phi}(a)$ is different from the efficient estimator when the function $\Pr[S=1 | X, \boldsymbol{W}]$ is not known. Specifically, when $\Pr[S = 1 | X, \boldsymbol{W}]$ is not known the efficient estimator would have the term $\big \{ 1-\widehat p(X_j, \boldsymbol{W}_j) \big\}$ replaced by the indicator for an observation belonging to the non-randomized clusters, $I(S_j = 0)$; see \textcite{dahabreh2020transportingStatMed}. Thus, at least in principle, knowledge of the sampling probability can lead to efficiency improvements. This phenomenon is analogous to well-known observations about the estimation of treatment effects in observational studies of point treatments with no unmeasured confounding. In that context, knowledge of the probability of treatment can be used to improve efficiency when estimating the average treatment effect on the treated, but not for the average treatment effect in the entire population underlying the observational study \parencite{hahn1998role}. In Appendix \ref{appendix:robustness}, we also show that $\widehat{\phi}(a)$ is robust in the sense that it converges to $\phi(a)$ whether or not the estimator $\widehat g_a(X, \boldsymbol{W})$ is consistent for $\E\left[ \overbar{Y} | X, \boldsymbol{W}, S = 1, A = a \right]$.

\paragraph{Inference:} We estimate the sampling variance of $\widehat \phi(a)$ as 
\begin{equation}
\widehat \sigma^2_{\widehat \phi(a)} = \dfrac{1}{m}  \widehat{Var} \Big[ \mathit{\widehat\Phi}^1_j(a) \Big],
\end{equation}
where $ \widehat{Var} \Big[ \mathit{\widehat{\Phi}}^1_j(a) \Big] $ is the sample variance of the influence curve $\mathit{\widehat\Phi}^1_j(a)$ (the ``sample analog'' of the influence function we give in Appendix \ref{appendix:influence functions}): 
\begin{equation*}
    \begin{split}
    \mathit{\widehat\Phi}^1_j(a) = \dfrac{1}{ \widehat \pi }\Big\{ \dfrac{I(S_j = 1, A_j = a)\big\{ 1- \widehat p(X_j, \boldsymbol{W}_j) \big\} }{\widehat p(X_j, \boldsymbol{W}_j) \widehat e_a(X_j, \boldsymbol{W}_j)} \Big\{ \overbar{Y}_j - \widehat g_a(X_j, \boldsymbol{W}_j) \Big\} + \\
    \big\{ 1-\widehat p(X_j, \boldsymbol{W}_j)\big\} \Big\{ \widehat g_a(X_j, \boldsymbol{W}_j) - \widehat \phi(a) \Big\} \Big\},
    \end{split}
\end{equation*} 
where $\widehat \pi$ is an estimator for $\Pr[S = 0]$, that is, $\widehat \pi = \frac{1}{m} \sum_{j = 1}^{m} I(S_j = 0)$.
The sampling variance can be used to obtain a $(1 - \alpha)\%$ confidence interval as $\Big( \widehat \phi(a) \pm z_{1-\alpha/2} \times \widehat \sigma_{\widehat \phi(a)} \Big),$ where $z_{1-\alpha/2}$ is the $(1-\alpha/2)$ quantile of the standard normal distribution. Inference may also be obtained via the non-parametric bootstrap \parencite{efron1994introduction}, accounting for clustering \parencite{balzer2019new}.

\subsection*{Average treatment effects}

The expectation of the average treatment effects can be estimated by taking differences between pairs of the estimators of the expectation of the average potential outcome described above. For example, the expectation of the average treatment effect in the entire target population, comparing treatments $a$ and $a^\prime$, using the augmented weighting estimator in equation \eqref{eq:AIPW_est_all}, can be estimated as $\widehat \psi(a) - \widehat \psi(a^\prime)$. Analogous treatment effect estimators can be obtained for the non-randomized subset of the target population.

\subsection*{Modeling participation, treatment, and outcomes} 

As noted above, the sampling probability and the probability of treatment in the trial are both known by design and can be used to estimate the expectation of the average potential outcomes and average treatment effects. Nevertheless, estimating these probabilities using simple parametric models (at the cluster level) can result in more precise estimates \parencite{lunceford2004, williamson2014variance}. We illustrate this behavior of the estimators in the simulation studies presented in the next section.

In contrast, the expectation of the average observed outcome conditional on baseline covariates and treatment in the trial, $\E\left[ \overbar{Y} | X, \boldsymbol{W}, S = 1, A = a \right]$, is not known and has to be estimated. To use individual-level information when estimating this expectation, we modify the strategy of \textcite{balzer2019new} for use in the context of analyses extending causal inferences to a target population. To begin, we specify and fit a working regression model for the conditional expectation of the individual-level outcome, $Y_{j,i}$, given cluster-level covariates, $X_j$, and the individual's covariates, $W_{j,i}$, (not the entire matrix $\boldsymbol{W}_{j}$); the regression can be fit separately by treatment arm, to allow for heterogeneity of treatment effects. Next, we obtain estimated values from the fitted model on all individuals in the data, regardless of trial selection status. We denote these predictions as $\widehat h_a(X_{j}, W_{j,i})$. Last, we obtain estimates $\widehat g_a(X_j, \boldsymbol{W}_j)$ by averaging the predictions over individuals in each cluster, $ \widehat g_a(X_j, \boldsymbol{W}_j) = \frac{1}{N_j} \sum_{i = 1}^{N_j} \widehat h_a(X_{j}, W_{j,i})$. We note that it is also possible (and may be necessary in some cases when data have already been aggregated to the cluster level) to estimate $\E\left[ \overbar{Y} | X, \boldsymbol{W}, S = 1, A = a \right]$ using only cluster-level data (e.g., by regressing cluster-level averages of the observed outcomes on cluster characteristics and cluster-level averages, or other summaries, of individual characteristics).

When individual-level data is available, the choice of whether to model the outcome at the individual-level or cluster-level is not obvious. In general, the model of the outcome conditional on covariates and treatment at the individual-level will be different than the model of the cluster-level average of the outcome. Furthermore, our ability to specify and estimate either of these models will depend on background knowledge, properties of the data generating mechanism, and data attributes (e.g., sample size, data availability, etc.). That said, we note that by using equation \eqref{eq:AIPW_est_all} or \eqref{eq:AIOW_est_S0}, our estimate remains consistent regardless of the specification of the outcome because the sampling probability is known by design and investigators can always correctly specify a model for it.

\section{Simulation studies} \label{sec:simulation}
We conducted a simulation study to verify the performance of the proposed augmented weighting estimators and to compare them against non-augmented weighting estimators. The choices in the simulation study, such as cluster size and the number of clusters in the sample from the target population, were informed by recently completed and ongoing trials of vaccine effectiveness in U.S. nursing homes \parencite{crt_trial1,crt_trial2,crt_trial3, gravenstein2016cluster, gravenstein2017comparative, gravenstein2021adjuvanted}. We considered scenarios with different treatment effects (including a scenario under the sharp null hypothesis), different magnitudes of heterogeneity (weaker or stronger), presence or absence of interference, and with or without outcome model misspecification (for models fit at the individual level). Table \ref{table_scenarios} summarizes the scenarios we considered; in the remainder of this section, we focus on describing Scenario 1 that has strong heterogeneity, presence of interference, and a correctly specified outcome model (at the individual level).

\paragraph{Baseline data generation:} We generated a sample of $m=5000$ trial-eligible clusters from the target population. Each cluster had a sample size $N_{j}$, $j = 1, \ldots, m$, randomly drawn from a Poisson distribution with mean parameter of 100. Thus, the number of individuals in each cluster varied, but on average there were approximately 100 individuals per cluster, which is similar to the sample sizes in the trials we used to motivate the simulation.

We generated a binary cluster-level covariate, $X_j$, with a Bernoulli distribution with parameter $\Pr[X_{j} = 1] = 0.05$. In each cluster, we generated individual-level covariates, by generating two column vectors of $\boldsymbol{W}_j$, that is, $\boldsymbol{W}_{1,j}$ and $\boldsymbol{W}_{2,j}$. For each cluster $j$, we denote the elements of $\boldsymbol{W}_{1,j}$ and $\boldsymbol{W}_{2,j}$ as $W_{1,j,i}$ and $W_{2,j,i}$, for $j = 1, \ldots, N_j$;  we generated these elements using draws from two independent cluster-specific normal distributions, each with its own mean and variance of 1. We independently drew the cluster-specific mean for each of these individual-level covariates from a continuous uniform distribution from -1 to 1. We define the cluster-level average of $W_1$ as $\overbar{W}_{1,j} = \frac{1}{N_j}\sum_{i = 1}^{N_j} W_{1,j,i}$ and the cluster-level average of $W_2$ as $\overbar{W}_{2,j}= \frac{1}{N_j}\sum_{i = 1}^{N_j} W_{2,j,i}.$

\paragraph{Selecting the clusters in the randomized trial with known sampling probabilities:} We simulated trials with different cluster sample sizes: 50, 100, or 200 clusters. We sampled clusters from a cohort of 5000 trial-eligible clusters into the randomized trial so that $\Pr[X_{j}=1 | S_j = 1]=0.5$; that is to say, we wanted the clusters enrolled in the trial to be (approximately) equally split between the two possible levels of the cluster-level covariate $X_j$. To accomplish this, we had to oversample clusters with $X_j = 1$ and undersample clusters with $X_j = 0$.  

For Bernoulli-type sampling of clusters from the target population sample, we used the sampling probability given by $$\Pr[S_j=1 | X_{j}=x]= \frac{\Pr[S_j = 1]}{\Pr[X_{j}=x]} \Pr[X_{j}=x|S_j=1], \mbox{ for } x = 0, 1.$$ For example, suppose that the targeted trial sample size was 50 clusters, the target population sample was 5000 clusters, and the desired proportion of clusters with $X_j = 1$ in the trial was 0.5. Then, among clusters with $X_j = 1$, we set the known-by-design sampling probability to $\Pr[S_j = 1 | X_{j} = 1]= \dfrac{50/5000}{0.05} 0.5 = 0.1$; similarly, among clusters with $X_j = 0$ we set the sampling probability to $\Pr[S_j = 1 | X_{j} = 0] \approx 0.005$. Note that when designing cluster randomized trials, the quantities $\Pr[S_j = 1]$ and $\Pr[X_{j}=x|S_j=1]$ would reflect the choice of the evaluators for the design of the trial and $\Pr[X_{j}=x]$ would be chosen on the basis of background knowledge about the target population or empirically estimated in the target population sample (in the simulation, we used the estimated $\Pr[X_{j}=x]$ value in each run of the simulation).

\paragraph{Treatment and outcome generation:} Treatment $A_j$ was randomized at the cluster-level, following a Bernoulli distribution with parameter $\Pr[A_j = 1 | S_j = 1] = 0.5$. For each individual $i$ in cluster $j$, we calculated the linear predictor $L_{j,i}= (2 A_j - 1) X_{j} + 0.5(2 A_j - 1) W_{1, j, i} + 0.5(2A_j-1) W_{2, j, i} +0.5(2A_j-1)\overbar{W}_{1,j}+0.5(2A_j-1)\overbar{W}_{2,j} $. We then simulated binary individual-level outcomes from a Bernoulli distribution with parameter $\Pr [ Y_{j,i} = 1|X_{j}, W_{1, j, i}, W_{2, j, i},  S_{j}=1,  A_{j} ] = \exp(L_{j,i} ) / \{ 1 + \exp(L_{j,i} ) \}$. Note that including $\overbar{W}_{1,j}$ and $\overbar{W}_{2,j}$ creates covariate interference. Also note that the product terms between the covariates and treatment in the generative model for the outcome induce heterogeneity of treatment effects.

\paragraph{Estimators:} We considered estimation of the expectation of average potential outcomes and the average treatment effects in the entire target population and its non-randomized subset. When estimating quantities in the entire target population, we applied the augmented inverse probability weighting estimator in equation \eqref{eq:AIPW_est_all} with the outcome-model fit using cluster-level information (AIPW1) or individual-level information (AIPW2). We also considered the following non-augmented inverse probability weighting estimator (IPW): 
\begin{equation}\label{eq:IPW_est_all}
    \begin{split}
    \widehat{\psi}_{\text{\tiny w}}(a) &= \dfrac{1}{m} \sum\limits_{j = 1}^{m} \dfrac{I(S_j = 1, A_j = a) \overbar{Y}_j}{\widehat p(X_j, \boldsymbol{W}_j) \widehat e_a(X_j, \boldsymbol{W}_j)}.
    \end{split}
\end{equation}
This estimator can be viewed as a special case of $\widehat\psi(a)$ with the outcome model terms $\widehat g_a(X, \boldsymbol{W})$ set identically to 0.

When estimating quantities in the non-randomized subset of the target population, we used the augmented inverse odds weighting estimator in equation \eqref{eq:AIOW_est_S0} with the outcome-model fit using only cluster-level information (AIOW1) or both cluster- and individual-level information (AIOW2). We compared these estimators against the following non-augmented inverse odds weighting estimator (IOW):  
\begin{equation}\label{eq:IOW_est_S0}
    \begin{split}
    \widehat{\phi}_{\text{\tiny w}}(a) &= \left\{\sum\limits_{j = 1}^{m} I(S_j = 0)\right\}^{-1} \sum\limits_{j = 1}^{m}   \dfrac{I(S_j = 1, A_j = a) \big\{ 1- \widehat p(X_j, \boldsymbol{W}_j) \big\} \overbar{Y}_j }{\widehat p(X_j, \boldsymbol{W}_j) \widehat e_a(X_j, \boldsymbol{W}_j)} .
    \end{split}
\end{equation}
Similar to the inverse probability weighting estimator above, $\widehat{\phi}_{\text{\tiny w}}(a)$ can be viewed as a special case of $\widehat\phi(a)$ with the outcome model terms $\widehat g_a(X, \boldsymbol{W})$ set identically to 0.

We note in passing that the augmented inverse probability weighting estimator (using the known-by-design or estimated probabilities) is asymptotically at least as efficient as the non-augmented inverse probability weighting estimator using the known-by-design probabilities, when the outcome model is correctly specified (see Appendix \ref{appendix:avar_comp}). Finally, we compared all these estimators against a trial-only estimator, which is estimated by averaging the individual-level outcomes in each cluster and then taking the average of these averages over the clusters participating in the trial.

\paragraph{Model specification:} For all estimators we considered three possible versions for the sampling probability and probability of treatment in the trial: one where the known-by-design probabilities were used, one where both probabilities were estimated with a simple logistic regression model at the cluster level (conditional only on the cluster-level variable, $X_j$, that determined the sampling probabilities), and one where both probabilities were estimated with a more complex logistic regression model at the cluster level (on the cluster-level variable, $X_j$, that determined the sampling probabilities and cluster-level averages of the individual-level covariates, $\boldsymbol{W}_j= (\boldsymbol{W}_{1,j}, \boldsymbol{W}_{2,j})$). 

For estimators that involve outcome modeling (i.e., AIPW and AIOW), we either modeled the outcome using cluster-level data (AIPW1 and AIOW1) or individual-level data (AIPW2 and AIOW2). When modeling using cluster-level data, we used a linear regression model for the cluster-specific average outcome, conditional on $X_j$ and the cluster-level averages of $\boldsymbol{W}_{1,j}$ and $\boldsymbol{W}_{2,j}$. When modeling using individual-level data, we used a logistic regression model for the indicator of the outcome, conditional on $X_j$ and the elements of $\boldsymbol{W}_{1,j}$ and $\boldsymbol{W}_{2,j}$ corresponding to each observation, along with the  cluster-level averages of $\boldsymbol{W}_{1,j}$ and $\boldsymbol{W}_{2,j}$, separately in each treatment arm. The outcome model was correctly specified when using individual-level data; the outcome model was misspecified when using cluster-level data (as would be the case in most practical applications, when the true underlying individual-level model is complex), and thus the regression of the cluster-level average of individual observed outcomes on cluster-level covariates is best viewed as an attempt to approximate the underlying true function.

\paragraph{Performance assessment:} We evaluated the performance of the estimators over 2000 simulation runs, in terms of bias and average standard deviation. We compared the estimated average standard deviation of the estimators (over the simulation runs) against (1) the average of the influence curve-based standard deviations, and (2) the average of a standard deviation estimated using a clustered bootstrap procedure that resamples with replacement from all the clusters \parencite{field2007bootstrapping}, with 500 bootstrap samples in each of the simulation runs. We also compared the coverage of the augmented weighting estimators when using the influence curve-based standard deviation versus the standard deviation estimated using the clustered bootstrap procedure. To facilitate numerical comparisons, we multiplied the simulation estimates of the bias and average standard deviation by the square root of the target cluster sample size, $\sqrt{5000}$. Because of the complexity of our data generating model, we obtained estimates of the ``true'' values for the expectations of the average potential outcomes and the average treatment effects using numerical methods (i.e., by generating both potential outcomes under the two levels of treatment for each observation over the simulation runs).  

\paragraph{Additional simulation scenarios (Scenarios 2 through 6):} Here, we briefly summarize the additional scenarios we examined in the simulation study (see also Table \ref{table_scenarios} for a summary of the different scenarios and Appendix \ref{appendix_addtional_sim_details} for additional details regarding the specification of models for data generation and for analyzing the data as needed for different estimators). Briefly, in Scenario 2 we modified Scenario 1 to generate data in the absence of interference; in Scenario 3 we considered weaker treatment effects and heterogeneity; and in Scenario 4 we generated data under the sharp null hypothesis of no treatment effect (and no effect heterogeneity). In Scenarios 5 and 6 we examined the impact of different kinds of misspecification of the outcome model. In Scenario 5, we generated data using the same approach as in Scenario 1, but when modeling the outcome at the individual-level we omitted $\boldsymbol{W}_{2,j}$ and its cluster-level average $\boldsymbol{W}_{2,j}$, and when modeling the outcome at the cluster level we omitted $\boldsymbol{W}_{2,j}$. In Scenario 6, we generated data with a nonlinear relationship (on the logit scale) between the outcome and $\boldsymbol{W}_{1,j}$, $\boldsymbol{W}_{2,j}$, and their corresponding cluster-level averages, but analyzed the data using models that only had linear terms. 

\paragraph{Results for Scenario 1:} Here, we present results from simulation Scenario 1 for estimands pertaining to the entire target population in the main text; results for estimands pertaining to the non-randomized subset of the target population were similar and are presented in Appendix \ref{appendix_sim_tables_non_randomized}. Table \ref{table_selected_results_simulation_targetALL_scenario1} shows the scaled bias (i.e., bias $\times \sqrt{5000}$) of each estimator across the simulation runs. The trial-only estimator, as expected, is biased for estimating the expectation of the average potential outcomes or the average treatment effect in the entire target population because sampling into the cluster randomized trial depends on the effect modifier, $X_j$. All estimators that account for the sampling of clusters for participation in the trial (IPW, AIPW1, AIPW2) show negligible bias, even in small cluster randomized trials.

In Table \ref{table_selected_results_simulation_targetALL_SD_scenario1}, we present the scaled standard deviation (i.e., the standard deviation over the simulation runs $\times \sqrt{5000}$) for estimators that account for the sampling of clusters for participation in the trial (IPW, AIPW1, AIPW2). When using the true sampling and treatment probabilities, IPW had substantially higher standard deviation compared with both AIPW1 and AIPW2. Estimating the sampling and treatment probabilities, using a simple model that only included $X_j$ (i.e., the variable used to determine the cluster sampling probability) or a more complex model that included $X_j$ and cluster-level averages of $\boldsymbol{W}_{1,j}$ and $\boldsymbol{W}_{2,j}$ reduced the standard deviation of IPW, but not enough to reach the standard deviation of AIPW1 or AIPW2. In contrast, AIPW1 and AIPW2 had similar performance when using the true sampling and treatment probabilities, when estimating these probabilities conditional on the variable $X_j$ (i.e., the variable used to determine the cluster sampling probability), or when estimating these probabilities using a more complex model that included $X_j$ and cluster-level averages that included $X_j$ and cluster-level averages of the elements of $\boldsymbol{W}_{1,j}$ and $\boldsymbol{W}_{2,j}$. 

To examine methods for statistical inference, we focus on the augmented weighting estimators (AIPW1 and AIPW2) because they were nearly unbiased and had a substantially lower standard deviation compared with IPW. Table \ref{table_selected_results_simulation_targetALL_coverage_scenario1} presents the average of the standard errors using the influence curve-based approach (IC) and the cluster bootstrap (BS), along with the corresponding coverage of Wald-style 95\% confidence intervals obtained using these standard errors for the augmented weighting estimators (AIPW1 and AIPW2). Ideally, the scaled average standard error should equal the scaled standard deviation in Table \ref{table_selected_results_simulation_targetALL_SD_scenario1}. In general, the influence curve-based approach for the average standard error was smaller than the standard deviation of the estimators, especially in smaller cluster trials. Using the influence curve-based standard errors resulted in undercoverage in the smaller cluster trials of 50 or 100 clusters, but nearly nominal coverage in larger cluster trials of 200 clusters. The average standard error based on the cluster bootstrap was similar to the standard deviation of the estimators. Using the bootstrap-based standard errors resulted in near-nominal coverage for all trial sizes we examined. Appendix \ref{appendix_sim_tables_non_randomized} summarizes results for Scenario 1 for estimators of average treatment effects and expectations of the average potential outcome in the non-randomized subset of the target population; the results were similar to the results reported here for the entire target population.

\paragraph{Results for additional simulation scenarios (2 through 6):} We report detailed results for these scenarios in Appendices \ref{appendix_scenario2} through  \ref{appendix_scenario6}. Regardless of the magnitude of the treatment effect and the amount of heterogeneity, in Scenarios 2 and 3, the trial-only estimator was biased, while the other estimators (IPW, AIPW1, and AIPW2) remained unbiased. Scenario 4, produced similar results for the IPW, AIPW1, and AIPW2 estimators (in this scenario, the trial-only estimator was also unbiased for the average treatment effect in the target population, but remained biased for expectations of the average potential outcome). In these scenarios, as in Scenario 1, the AIPW estimators had smaller standard deviation than IPW. In Scenarios 5 and 6, where we examined the impact of misspecifying the outcome model, the AIPW estimators had little bias because regardless of the specification of the outcome model, the probability of participation was either the true one (known by design) or estimated using a correctly specified model (i.e., the simulations reflect the robustness property of the AIPW estimator).

\section{Discussion}

We described a nested cluster randomized trial design where clusters are selected for inclusion in the trial with known sampling probabilities that depend on baseline covariates, and proposed robust augmented weighting estimators for this design. The robustness of the proposed estimators stems from the fact that the sampling probability and the probability of treatment in the trial are known by design and thus models for them can always be correctly specified. Our estimators give evaluators the option of exploiting individual-level data on the relationship between covariates, treatment and outcomes, to further increase efficiency, while accounting for within-cluster dependence, including various forms of interference. We showed that, for causal estimands that pertain to the non-randomized subset of the target population, knowledge of the sampling probabilities can be used to develop augmented weighting estimators that are more efficient compared to augmented weighting estimators when the sampling probabilities are unknown. This improvement did not occur for estimands that pertain to the entire target population because their efficient influence function is the same, whether the sampling probabilities are known or unknown.

Our proof-of-concept simulations, motivated by large cluster randomized trials of vaccine effectiveness \parencite{crt_trial1,crt_trial2,crt_trial3}, show that the augmented weighting estimators perform well in finite samples and better than previously described non-augmented weighting estimators. The augmented weighting estimators had about the same performance whether the true or estimated sampling and treatment probabilities were used, even in small trials. In contrast, our simulation results suggest that estimating the known-by-design sampling and treatment probabilities when using the non-augmented weighting estimators can substantially improve precision, but the improvement is often not enough to reach the precision of the augmented weighting estimators. In the simulation, the standard deviation estimated using a clustered bootstrap procedure worked well for inference with the augmented weighting estimators and the influence curve-based standard deviation (which is computationally faster) also performed well in larger cluster trial sizes. Even though the clustered bootstrap procedure we used worked well in our simulations, there are many options for bootstrapping clustered data \parencite{field2007bootstrapping, davison1997bootstrap} and comparisons among them might be useful, particularly for studies with a smaller number of clusters.

Prior work on designing a cluster randomized trial to support generalizable inferences has focused on sampling clusters such that crude (unadjusted) analyses of the trial data can estimate treatment effects in the target population \parencite{tipton2013stratified, tipton2014sample, tipton2017design}. Such ``representative'' sampling using a constant sampling probability across strata defined by effect modifiers puts a premium on being able to use relatively simple statistical analyses but cannot accommodate other practical aspects of trial conduct, such as the need for rapid recruitment of clusters, the recruitment of clusters with established research infrastructure, or the desire for efficient estimation within subgroups of clusters defined by covariates. When representative sampling does not result in good balance between the trial and the target population, this prior work has mentioned the possibility of using simple weighting or stratification methods, without providing evidence of good performance. Our proposed design essentially works in the opposite direction, by acknowledging that evaluators often have to select clusters for participation conditional on baseline covariates, while still wanting to draw generalizable inferences -- an instance of experimental design with multiple objectives \parencite{sverdlov2013recent, woodcock2017master, sverdlov2020optimal}. For example, our approach can support trials designed to test hypotheses in subgroups of clusters, by oversampling clusters with certain characteristics, and uses the known-by-design sampling probabilities to produce inferences that apply to the target population.

Our proposed design and analysis methods should be useful when practicalities of trial conduct (e.g., efficient recruitment) or the trial's inferential goals require oversampling clusters with certain characteristics. They can form the basis for explicit approaches to the planning of future cluster randomized trials \parencite{raudenbush1997statistical,copas2021optimal} via formal optimization procedures for trading-off competing research objectives. Such optimization efforts are motivated by the desire to use the most efficient experimental designs that are feasible; thus, they should routinely be paired with efficient estimation approaches, such as those that we have proposed.

In most cases evaluators will choose sampling probabilities that depend on a low dimensional set of discrete covariates (e.g., those deemed as the most likely and strong effect modifiers for the treatment effects of interest). That said, our methods can also accommodate more complex sampling schemes. For example, when evaluators would like to sample clusters based on multiple covariates (as may be the case in large cluster randomized trials that motivate our work \parencite{crt_trial1,crt_trial2,crt_trial3}), a risk or effect score, or other dimensionality reduction approach, could be used to create a lower dimensional variable that can then be used to determine the sampling probabilities. 

Throughout, our exposition assumed that the evaluators have complete control over cluster participation in the trial. Nevertheless, the methods can be easily extended to allow for the possibility that selected clusters may decline participation in the trial. When sampled clusters can decline participation, additional causal assumptions regarding the exchangeability of clusters that agree to participate with those that do not, among the sampled clusters, will be needed (this is analogous to recent results for individually randomized trials \parencite{dahabreh2019identification}). Furthermore, the model for the probability of participation among sampled clusters will need to be correctly specified on the basis of background knowledge, because participation among the sampled clusters will not be under the control of the evaluators.

In summary, cluster randomized trials where clusters are selected for inclusion with known sampling probabilities that depend on cluster characteristics, combined with efficient estimation methods, can lead to substantial improvements in the precision of the estimated effect in the target population, while also addressing competing objectives of trial conduct.

\section{Acknowledgment}

This work was supported in part by National Library of Medicine (NLM) Award R01LM013616, and Patient-Centered Outcomes Research Institute (PCORI) awards ME-1502-27794 and ME-2019C3-17875.
The content of this paper is solely the responsibility of the authors and does not necessarily represent the official views of the NLM, PCORI, the PCORI Board of Governors, or the PCORI Methodology Committee.

\section{Declaration}

Dr. Dahabreh is the principal investigator of a research agreement between Harvard University and Sanofi on transportability methods for individually randomized trials, unrelated to this manuscript; Dr. Dahabreh also reports consulting fees from Moderna for work unrelated to this manuscript. The other authors report no potential conflicts of interest.

\clearpage
\printbibliography


\ddmmyyyydate 
\newtimeformat{24h60m60s}{\twodigit{\THEHOUR}.\twodigit{\THEMINUTE}.32}
\settimeformat{24h60m60s}
\begin{center}
\vspace{\fill}\ \newline
\textcolor{black}{{\tiny $ $generalizability\_clusters\_designed, $ $ }
{\tiny $ $Date: \today~~ \currenttime $ $ }
{\tiny $ $Revision: \paperversionmajor.\paperversionminor $ $ }}
\end{center}

\clearpage
\section*{Tables}
\begin{table}[ht!]
\centering
\caption{Simulation study scenarios for the outcome data generating mechanism and outcome model specification.}
\label{table_scenarios}
\begin{tabular}{@{}ccccc@{}}
\toprule
Scenario                                 &  \begin{tabular}[c]{@{}l@{}}Treatment effect \& \\ heterogeneity\end{tabular}  & Interference & Linear & Outcome model \\ \midrule
1                                    & Stronger                                                           & Yes           & Yes    & Correctly specified                                                                     \\
2                                   & Stronger                                                           & No           & Yes    & Correctly specified                                                                     \\
3                                  & Weaker                                                           & Yes           & Yes    & Correctly specified                                                                     \\
4                                  & None (sharp null)                                                          & Yes           & Yes    & Correctly specified                                                                     \\
5                                 & Stronger                                                          & Yes           & Yes    & Misspecified                                                                    \\
6                                 & Stronger                                                          & Yes           & No    & Misspecified                                                                                                                                \\ \bottomrule
\end{tabular}
\caption*{Scenario 1 is the scenario presented in the main text; Scenario 2 does not have interference; Scenario 3 considers a weaker treatment effect and weaker heterogeneity; Scenario 4 has no treatment effect (and no heterogeneity); Scenarios 5 and 6 examine the impact of outcome model misspecification. \\

Effect heterogeneity = level of heterogeneity, where none (sharp null) means no treatment effect or heterogeneity in the outcome generating mechanism; Interference = indicates whether covariate interference is present in the outcome data generating mechanism; Linear = indicates whether nonlinear terms are present in the outcome data generating mechanism; Outcome model = indicates whether the individual-level outcome model is correctly specified.}
\end{table}

\renewcommand{\arraystretch}{1.16}
\begin{table}[ht!]
\centering
\caption{Scaled bias of estimators for quantities in the target population.}
\label{table_selected_results_simulation_targetALL_scenario1}
\small
\begin{tabular}{lcccccc}
\toprule
Estimand & $n$   & Values for probabilities & Trial-only  & IPW     & AIPW1   & AIPW2   \\ \midrule
ATE          & 50                        & True                & 12.286 & -0.045 & 0.025  & 0.025  \\
         & 50                        & Estimated (simple)  & 12.286 & 0.015  & 0.025  & 0.025  \\
          & 50                        & Estimated (complex) & 12.286 & 0.001  & -0.019 & 0.025  \\
          & 100                       & True                & 12.149 & -0.232 & 0.01   & 0.014  \\
         & 100                       & Estimated (simple)  & 12.149 & -0.048 & 0.01   & 0.013  \\
         & 100                       & Estimated (complex) & 12.149 & 0.047  & -0.005 & 0.015  \\
         & 200                       & True                & 12.296 & 0.127  & 0.000  & 0.000  \\
         & 200                       & Estimated (simple)  & 12.296 & 0.012  & 0.000  & 0.000  \\
          & 200                       & Estimated (complex) & 12.296 & -0.026 & -0.012 & 0.000  \\ \midrule
$\E \big [\overbar{Y}^{a=1} \big]$           & 50                        & True                & 6.099  & -0.144 & 0.013  & 0.014  \\
         & 50                        & Estimated (simple)  & 6.099  & -0.024 & 0.013  & 0.014  \\
          & 50                        & Estimated (complex) & 6.099  & 0.040  & -0.005 & 0.011  \\
        & 100                       & True                & 6.062  & -0.204 & 0.001  & 0.007  \\
         & 100                       & Estimated (simple)  & 6.062  & -0.060 & 0.001  & 0.006  \\
         & 100                       & Estimated (complex) & 6.062  & -0.010 & -0.007 & 0.008  \\
          & 200                       & True                & 6.142  & 0.012  & -0.005 & -0.006 \\
          & 200                       & Estimated (simple)  & 6.142  & -0.014 & -0.005 & -0.006 \\
          & 200                       & Estimated (complex) & 6.142  & -0.035 & -0.01  & -0.004 \\ \midrule
$\E \big [\overbar{Y}^{a=0} \big]$        & 50                        & True                & -6.187 & -0.100 & -0.011 & -0.011 \\
       & 50                        & Estimated (simple)  & -6.187 & -0.039 & -0.011 & -0.012 \\
          & 50                        & Estimated (complex) & -6.187 & 0.039  & 0.014  & -0.014 \\
         & 100                       & True                & -6.088 & 0.027  & -0.009 & -0.007 \\
       & 100                       & Estimated (simple)  & -6.088 & -0.011 & -0.009 & -0.007 \\
         & 100                       & Estimated (complex) & -6.088 & -0.057 & -0.001 & -0.008 \\
         & 200                       & True                & -6.153 & -0.114 & -0.005 & -0.006 \\
        & 200                       & Estimated (simple)  & -6.153 & -0.026 & -0.005 & -0.006 \\
         & 200                       & Estimated (complex) & -6.153 & -0.009 & 0.002  & -0.004  \\  \bottomrule
\end{tabular}
\caption*{Results are scaled by $\sqrt{m}$ (i.e., multiplied by $\sqrt{5000} \approx 70.7$). ATE is defined as $\E \big [\overbar{Y}^{a=1} \big] - \E \big [\overbar{Y}^{a=0} \big] $; $n$ = number of clusters in the trial; Values for probabilities = how the treatment and sampling probabilities are obtained for the estimators (True = use the known-by-design sampling probabilities and probabilities of treatment in the trial; Estimated (simple) = estimate the sampling and treatment probabilities conditional on $X_j$ only (the variable used to determine the sampling probabilities); Estimated (complex) = estimate the sampling and treatment probabilities conditional on $X_j$ and the cluster-level averages of $\boldsymbol{W}_{1,j}$ and $\boldsymbol{W}_{2,j}$); Trial-only = average the individual-level outcomes in each cluster and then take the average of these averages over the clusters participating in the trial; IPW = non-augmented inverse probability weighting estimator; AIPW1 = augmented inverse probability weighting estimator, with the outcome model fit only at the cluster-level; AIPW2 = augmented inverse probability weighting estimator, with the outcome model fit at the individual-level.}
\end{table}
\renewcommand{\arraystretch}{1}

\renewcommand{\arraystretch}{1.16}
\begin{table}[ht!]
\centering
\caption{Scaled standard deviation of estimators for quantities in the target population.}
\label{table_selected_results_simulation_targetALL_SD_scenario1}
\begin{tabular}{lccccc}
\toprule
Estimand & $n$   & Values for probabilities & IPW & AIPW1 & AIPW2 \\ \midrule
ATE      & 50                        & True                & 14.133 & 1.453    & 1.384    \\
    & 50                        & Estimated (simple)  & 4.847  & 1.453    & 1.385    \\
     & 50                        & Estimated (complex) & 5.123  & 1.484    & 1.427    \\
     & 100                       & True                & 10.144 & 0.989    & 0.944    \\
      & 100                       & Estimated (simple)  & 3.395  & 0.989    & 0.944    \\
      & 100                       & Estimated (complex) & 2.812  & 0.992    & 0.955    \\
      & 200                       & True                & 7.197  & 0.738    & 0.710    \\
     & 200                       & Estimated (simple)  & 2.308  & 0.738    & 0.710    \\
     & 200                       & Estimated (complex) & 1.649  & 0.733    & 0.714    \\ \midrule
$\E \big [\overbar{Y}^{a=1} \big]$      & 50                        & True                & 9.885  & 1.050    & 0.988    \\
     & 50                        & Estimated (simple)  & 3.464  & 1.050    & 0.989    \\
      & 50                        & Estimated (complex) & 3.783  & 1.072    & 1.019    \\
    & 100                       & True                & 7.073  & 0.671    & 0.637    \\
     & 100                       & Estimated (simple)  & 2.424  & 0.671    & 0.637    \\
     & 100                       & Estimated (complex) & 2.081  & 0.675    & 0.647    \\
      & 200                       & True                & 5.094  & 0.504    & 0.479    \\
     & 200                       & Estimated (simple)  & 1.688  & 0.504    & 0.479    \\
      & 200                       & Estimated (complex) & 1.236  & 0.497    & 0.481    \\ \midrule
$\E \big [\overbar{Y}^{a=0} \big]$      & 50                        & True                & 10.088 & 0.984    & 0.937    \\
      & 50                        & Estimated (simple)  & 3.444  & 0.984    & 0.938    \\
      & 50                        & Estimated (complex) & 3.803  & 0.992    & 0.960    \\
      & 100                       & True                & 7.077  & 0.693    & 0.650    \\
      & 100                       & Estimated (simple)  & 2.479  & 0.693    & 0.650    \\
      & 100                       & Estimated (complex) & 2.061  & 0.689    & 0.657    \\
      & 200                       & True                & 5.019  & 0.486    & 0.467    \\
      & 200                       & Estimated (simple)  & 1.650  & 0.486    & 0.467    \\
      & 200                       & Estimated (complex) & 1.238  & 0.482    & 0.468    \\
        \bottomrule
\end{tabular}
\caption*{Results are scaled by $\sqrt{m}$ (i.e., multiplied by $\sqrt{5000} \approx 70.7$). ATE is defined as $\E \big [\overbar{Y}^{a=1} \big] - \E \big [\overbar{Y}^{a=0} \big] $; $n$ = number of clusters in the trial; Values for probabilities = how the treatment and sampling probabilities are obtained for the estimators; True = use the known-by-design sampling probabilities and probabilities of treatment in the trial; Estimated (simple) = estimate the sampling and treatment probabilities conditional on $X_j$ only (the variable used to determine the sampling probabilities); Estimated (complex) = estimate the sampling and treatment probabilities conditional on $X_j$ and the cluster-level averages of $\boldsymbol{W}_{1,j}$ and $\boldsymbol{W}_{2,j}$; IPW = non-augmented inverse probability weighting estimator; AIPW1 = augmented inverse probability weighting estimator, with the outcome model fit only at the cluster-level; AIPW2 = augmented inverse probability weighting estimator, with the outcome model fit at the individual-level.}
\end{table}
\renewcommand{\arraystretch}{1}

\renewcommand{\arraystretch}{1.16}
\begin{table}[ht!]
\centering
\caption{Coverage and scaled average of standard errors of the augmented weighting estimators for quantities in the target population.}
\label{table_selected_results_simulation_targetALL_coverage_scenario1}
\resizebox{\textwidth}{!}{

\begin{tabular}{lcccccccccc@{}}
\toprule
\multicolumn{1}{l}{}               & \multicolumn{1}{c}{}    & \multicolumn{1}{c}{}     & \multicolumn{4}{c}{AIPW1}                                                                         & \multicolumn{4}{c}{AIPW2}   \\ [0.1cm]
\multicolumn{1}{l}{Estimand}       & \multicolumn{1}{c}{$n$} & \multicolumn{1}{c}{ Values for probabilities} & \multicolumn{2}{c}{ASE}                          & \multicolumn{2}{c}{Coverage}                    & \multicolumn{2}{c}{ASE}                          & \multicolumn{2}{c}{Coverage}                    \\ [0.1cm]
\multicolumn{1}{l}{}               & \multicolumn{1}{c}{}    & \multicolumn{1}{c}{}     & \multicolumn{1}{c}{IC} & \multicolumn{1}{c}{BS} & \multicolumn{1}{c}{IC} & \multicolumn{1}{c}{BS} & \multicolumn{1}{c}{IC} & \multicolumn{1}{c}{BS} & \multicolumn{1}{c}{IC} & \multicolumn{1}{c}{BS} \\ \midrule
ATE      & 50  & True                & 1.205         & 1.778         & 0.878          & 0.969          & 1.144         & 1.518         & 0.884          & 0.959          \\
     & 50  & Estimated (simple)  & 1.246         & 1.778         & 0.907          & 0.969          & 1.177         & 1.518         & 0.902          & 0.959          \\
      & 50  & Estimated (complex) & 1.292         & 1.911         & 0.904          & 0.963          & 1.210         & 1.625         & 0.899          & 0.960          \\
      & 100 & True                & 0.927         & 1.022         & 0.926          & 0.951          & 0.885         & 0.965         & 0.931          & 0.950          \\
      & 100 & Estimated (simple)  & 0.944         & 1.022         & 0.936          & 0.951          & 0.899         & 0.964         & 0.936          & 0.952          \\
     & 100 & Estimated (complex) & 0.964         & 1.022         & 0.938          & 0.954          & 0.915         & 0.976         & 0.936          & 0.956          \\
      & 200 & True                & 0.713         & 0.742         & 0.933          & 0.945          & 0.686         & 0.709         & 0.933          & 0.947          \\
      & 200 & Estimated (simple)  & 0.72          & 0.742         & 0.942          & 0.945          & 0.691         & 0.709         & 0.941          & 0.946          \\
     & 200 & Estimated (complex) & 0.727         & 0.734         & 0.942          & 0.941          & 0.697         & 0.712         & 0.944          & 0.948          \\ \midrule
$\E \big [\overbar{Y}^{a=1} \big]$      & 50  & True                & 0.822         & 1.189         & 0.851          & 0.951          & 0.777         & 1.034         & 0.853          & 0.944          \\
     & 50  & Estimated (simple)  & 0.853         & 1.189         & 0.879          & 0.951          & 0.803         & 1.034         & 0.885          & 0.944          \\
      & 50  & Estimated (complex) & 0.881         & 1.262         & 0.882          & 0.959          & 0.824         & 1.095         & 0.879          & 0.946          \\
      & 100 & True                & 0.627         & 0.697         & 0.913          & 0.952          & 0.595         & 0.655         & 0.915          & 0.949          \\
      & 100 & Estimated (simple)  & 0.642         & 0.697         & 0.937          & 0.952          & 0.608         & 0.655         & 0.929          & 0.948          \\
      & 100 & Estimated (complex) & 0.656         & 0.699         & 0.934          & 0.949          & 0.620         & 0.663         & 0.927          & 0.946          \\
     & 200 & True                & 0.473         & 0.493         & 0.924          & 0.939          & 0.452         & 0.468         & 0.928          & 0.944          \\
     & 200 & Estimated (simple)  & 0.476         & 0.493         & 0.934          & 0.939          & 0.455         & 0.468         & 0.935          & 0.945          \\
     & 200 & Estimated (complex) & 0.482         & 0.489         & 0.938          & 0.938          & 0.459         & 0.471         & 0.937          & 0.942          \\ \midrule
$\E \big [\overbar{Y}^{a=0} \big]$       & 50  & True                & 0.819         & 1.186         & 0.866          & 0.950          & 0.774         & 1.033         & 0.861          & 0.951          \\
      & 50  & Estimated (simple)  & 0.848         & 1.186         & 0.891          & 0.950          & 0.799         & 1.033         & 0.892          & 0.951          \\
      & 50  & Estimated (complex) & 0.877         & 1.276         & 0.893          & 0.955          & 0.820         & 1.105         & 0.889          & 0.952          \\
     & 100 & True                & 0.629         & 0.695         & 0.902          & 0.943          & 0.598         & 0.654         & 0.913          & 0.943          \\
      & 100 & Estimated (simple)  & 0.639         & 0.695         & 0.919          & 0.943          & 0.607         & 0.654         & 0.923          & 0.944          \\
      & 100 & Estimated (complex) & 0.652         & 0.694         & 0.927          & 0.941          & 0.617         & 0.661         & 0.929          & 0.942          \\
      & 200 & True                & 0.472         & 0.494         & 0.932          & 0.944          & 0.451         & 0.469         & 0.930          & 0.944          \\
      & 200 & Estimated (simple)  & 0.478         & 0.494         & 0.934          & 0.944          & 0.456         & 0.469         & 0.939          & 0.944          \\
      & 200 & Estimated (complex) & 0.483         & 0.489         & 0.937          & 0.938          & 0.460         & 0.471         & 0.938          & 0.942                \\ \bottomrule
\end{tabular}

}
\caption*{
ASE = average (over the simulations) of standard errors, scaled by $\sqrt{m}$ (i.e., multiplied by $\sqrt{5000} \approx 70.7$); IC = influence-curve based; BS= bootstrap; Coverage = coverage, using 95\% normal confidence intervals. ATE is defined as $\E \big [\overbar{Y}^{a=1} \big] - \E \big [\overbar{Y}^{a=0} \big] $; $n$ = number of clusters in the trial; Values for probabilities = how the treatment and sampling probabilities are obtained for the estimators; True = use the known-by-design sampling probabilities and probabilities of treatment in the trial; Estimated (simple) = estimate the sampling and treatment probabilities conditional on $X_j$ only (the variable used to determine the sampling probabilities); Estimated (complex) = estimate the sampling and treatment probabilities conditional on $X_j$ and the cluster-level averages of $\boldsymbol{W}_{1,j}$ and $\boldsymbol{W}_{2,j}$; AIPW1 = augmented inverse probability weighting estimator, with the outcome model fit only at the cluster-level; AIPW2 = augmented inverse probability weighting estimator, with the outcome model fit at the individual-level.}
\end{table}
\renewcommand{\arraystretch}{1}

\clearpage
\setcounter{page}{1}
\appendix

\section{Identification results under a nonparametric model}\label{appendix:identification}

This section of the Appendix summarizes results from a technical report \parencite{dahabreh2022clusterefficiency} dealing with cluster randomized trials nested within cohorts of trial eligible individuals, under a non-parametric model in which \emph{the sampling probability is not under the control of the evaluators}. These results remain valid when the sampling probabilities are known and are presented here for completeness.

\paragraph{Expectation of the average potential outcome in the target population:} Under identifiability conditions A1 through A5, given in the main text, the expectation of the average potential outcome in the target population, $ \E\left[\overbar{Y}^{a}\right] $, is identified by $$\psi(a) \equiv \E \Big[ \E\left[\overbar{Y} | X, \boldsymbol{W}, S = 1, A = a \right] \Big].$$

Starting with the causal quantity of interest,
\begin{equation*}
    \begin{split}
        \E\left[\overbar{Y}^{a}\right] &= \E \Big[ \E\left[\overbar{Y}^{a} | X, \boldsymbol{W} \right] \Big] \\
            &= \E \Big[ \E\left[\overbar{Y}^{a} | X, \boldsymbol{W}, S = 1 \right] \Big] \\
            &= \E \Big[ \E\left[\overbar{Y}^{a} | X, \boldsymbol{W}, S = 1, A = a \right] \Big] \\
            &= \E \Big[ \E\left[\overbar{Y} | X, \boldsymbol{W}, S = 1, A = a \right] \Big] \\
            &\equiv \psi(a),
    \end{split}
\end{equation*}
where the first step follows from the law of total expectation, the second by condition \emph{A4}, the third by condition \emph{A2}, the fourth by condition \emph{A1}, the last by the definition of $\psi(a)$, and quantities are well-defined because of the positivity conditions \emph{A3} and \emph{A5}. 

\paragraph{Expectation of the average potential outcome in the non-randomized subset of the target population:} Under identifiability conditions A1 through A4 and condition A5$^*$, given in the main text, the expectation of the average potential outcome in the non-randomized subset of the target population, $ \E\left[\overbar{Y}^{a} | S  = 0 \right] $, is identified by $$\phi(a) \equiv \E \Big[ \E\left[\overbar{Y} | X, \boldsymbol{W}, S = 1, A = a \right] \big| S = 0 \Big].$$

Starting with the causal quantity of interest,
\begin{equation*}
    \begin{split}
        \E\left[\overbar{Y}^{a}\right  | S = 0  ] &= \E \Big[ \E\left[\overbar{Y}^{a} | X, \boldsymbol{W} , S = 0 \right]  \big| S = 0  \Big] \\
            &= \E \Big[ \E\left[\overbar{Y}^{a} | X, \boldsymbol{W}, S = 1 \right] \big| S = 0 \Big] \\
            &= \E \Big[ \E\left[\overbar{Y}^{a} | X, \boldsymbol{W}, S = 1, A = a \right] \big| S = 0 \Big] \\
            &= \E \Big[ \E\left[\overbar{Y} | X, \boldsymbol{W}, S = 1, A = a \right] \big| S = 0 \Big] \\
            &\equiv \phi(a),
    \end{split}
\end{equation*}
where the first step follows from the law of total expectation conditional on $S = 0$, the second by condition \emph{A4}, the third by condition \emph{A2}, the fourth by condition \emph{A1}, the last by the definition of $\phi(a)$, and quantities are well-defined because of the positivity conditions \emph{A3} and \emph{A5$^*$}. 

\clearpage
\section{Estimation under the semiparametric model where the sampling probability and the probability of treatment are known}\label{appendix:influence functions}

\paragraph{Expectation of the average potential outcome in the target population:} In a technical report \parencite{dahabreh2022clusterefficiency}, we have shown that the influence function of $\psi(a)$ under a non-parametric model for the law of the observed data is  
\begin{equation*}
    \begin{split}
    \mathit{\Psi}^1_{p_0}(a) &= \dfrac{I(S = 1, A = a)}{\Pr_{p_0}[S = 1 | X, \boldsymbol{W}] \Pr_{p_0}[A  = a | X, \boldsymbol{W}, S = 1]} \Big\{ \overbar{Y} - \E_{p_0}\left[ \overbar{Y} | X, \boldsymbol{W}, S = 1, A = a \right] \Big\} \\
    &\quad\quad\quad\quad\quad+ \E_{p_0}\left[ \overbar{Y} | X, \boldsymbol{W}, S = 1, A = a \right] - \psi_{p_0}(a),
    \end{split}
\end{equation*}
where the subscript $p_0$ indicates the ``true'' law; $\Pr_{p_0}[S = 1 | X, \boldsymbol{W}]$ is the cluster-level probability of participating in the trial given $(X, \boldsymbol{W})$; $\Pr_{p_0}[ A  = a | X, \boldsymbol{W}, S = 1]$ is the probability of being assigned to treatment $a$ given $(X, \boldsymbol{W})$ among clusters participating in the trial; and $\E_{p_0}\left[ \overbar{Y} | X, \boldsymbol{W}, S = 1, A = a \right]$ is the conditional expectation of the cluster-level average observed outcome $ \overbar{Y} $ given $(X, \boldsymbol{W})$ among clusters participating in the trial and assigned to treatment $a$.

Now consider the semiparametric model with $\Pr[S = 1 | X, \boldsymbol{W}]$ known to the evaluators and the tangent space of that model. The influence function $\mathit{\Psi}^1_{p_0}(a)$ above belongs to the tangent space of this semiparametric model and thus $\mathit{\Psi}^1_{p_0}(a)$  is equal to its own projection onto the tangent space of the semiparametric model. It follows that $\mathit{\Psi}^1_{p_0}(a)$  is also the influence function under the semiparametric model with $\Pr[S = 1 | X, \boldsymbol{W}]$ known. Furthermore, a similar argument holds for the semiparametric model with both  $\Pr[S = 1 | X, \boldsymbol{W}]$ and $\Pr[ A  = a | X, \boldsymbol{W}, S]$ known. 

Thus, we can conclude that $\widehat \psi(a)$, that is, the estimating equation estimator based on the influence function $\mathit{\Psi}^1_{p_0}(a)$, is an efficient estimator under the semiparametric models with $\Pr[S = 1 | X, \boldsymbol{W}]$, or both $\Pr[S = 1 | X, \boldsymbol{W}]$ and $\Pr[ A  = a | X, \boldsymbol{W}, S]$, known.

\paragraph{Expectation of the average potential outcome in the non-randomized subset of the target population:} 
The influence function of $\phi(a)$ \cite{dahabreh2022clusterefficiency} under a non-parametric model for the law of the observed data is 
\begin{equation*}
    \begin{split}
    \mathit{\Phi}^1_{p_0}(a) &= \dfrac{1}{\pi_{p_0}} \Bigg\{ \dfrac{I(S = 1, A = a) \big\{ 1 - \Pr_{p_0}[S = 1 | X, \boldsymbol{W}] \big\} }{\Pr_{p_0}[S = 1 | X, \boldsymbol{W}] \Pr_{p_0}[A  = a | X, \boldsymbol{W}, S = 1]} \Big\{ \overbar{Y} - \E_{p_0}\left[ \overbar{Y} | X, \boldsymbol{W}, S = 1, A = a \right] \Big\} \\
    &\quad\quad\quad\quad\quad + I(S = 0) \big\{ \E_{p_0}\left[ \overbar{Y} | X, \boldsymbol{W}, S = 1, A = a \right] - \phi_{p_0}(a) \big\} \Bigg\},
    \end{split}
\end{equation*}
where $\pi_{p_0} = \Pr_{p_0}[S = 0]$. 

Now consider a semiparametric model with $\Pr[S = 1 | X, \boldsymbol{W}]$ known. To obtain the efficient influence function under this semiparametric model, we need to project $\mathit{\Psi}^1_{p_0}(a)$ to the tangent space of this semiparametric model. Using Theorems 4.3 and 4.5 of \textcite{tsiatis2007} and standard iterated expectation arguments, we find that the efficient influence function under this semiparametric model is
\begin{equation*}
    \begin{split}
    \widetilde{\mathit{\Phi}}^1_{p_0}(a) &= \dfrac{1}{\pi_{p_0}} \Bigg\{    \dfrac{I(S = 1, A = a) \big\{ 1 - \Pr_{p_0}[S = 1 | X, \boldsymbol{W}] \big\} }{\Pr_{p_0}[S = 1 | X, \boldsymbol{W}] \Pr_{p_0}[A  = a | X, \boldsymbol{W}, S = 1]} \Big\{ \overbar{Y} - \E_{p_0}\left[ \overbar{Y} | X, \boldsymbol{W}, S = 1, A = a \right] \Big\} \\
    &\quad\quad\quad\quad + \big\{ 1- \text{Pr}_{p_0}[S = 1 | X, \boldsymbol{W}] \big\} \big\{ \E_{p_0}\left[ \overbar{Y} | X, \boldsymbol{W}, S = 1, A = a \right] - \phi_{p_0}(a) \big\}  \Bigg\}.
    \end{split}
\end{equation*}
The difference between $\mathit{\Phi}^1_{p_0}(a)$ and $\widetilde{\mathit{\Phi}}^1_{p_0}(a)$ is the substitution of $\big\{ 1 - \Pr_{p_0}[S = 1 | X, \boldsymbol{W}] \big\}$ for $I(S = 0)$ in the second term of the latter. 

Furthermore, an argument similar to the one presented above for $\mathit{\Psi}^1_{p_0}(a)$ can be used to show that $\widetilde{\mathit{\Phi}}^1_{p_0}(a)$ is the efficient influence function in the semiparametric model with both $\Pr[S = 1 | X, \boldsymbol{W}]$ and $\Pr[ A  = a | X, \boldsymbol{W}, S]$ known. Thus, we can conclude that $\widehat \phi(a)$, that is, the estimating equation estimator based on the above efficient influence function $\widetilde{\mathit{\Phi}}^1_{p_0}(a)$, is an efficient estimator under the semiparametric models with $\Pr[S = 1 | X, \boldsymbol{W}]$, or both $\Pr[S = 1 | X, \boldsymbol{W}]$ and $\Pr[ A  = a | X, \boldsymbol{W}, S]$, known. 

For completeness, in Appendix \ref{appendix:avar_comp} we shall compare the asymptotic variance of the augmented inverse odds weighting estimator (corresponding to the influence function, $\widetilde{\mathit{\Phi}}^1_{p_0}(a)$) versus the asymptotic variance of the augmented inverse odds weighting estimator in the technical report (corresponding to the influence function, $\mathit{\Phi}^1_{p_0}(a)$). This shows that the asymptotic variance of the augmented odds weighting estimator corresponding to  $\widetilde{\mathit{\Phi}}^1_{p_0}(a)$ is less than or equal to the estimator corresponding to $\mathit{\Phi}^1_{p_0}(a).$

\clearpage 
\section{Robustness}\label{appendix:robustness}

In this section, we use $\overset{p}{\rightarrow}$ to denote convergence in probability.

\subsection{Expectation of the average potential outcome in the target population}

Recall that the estimator of the expectation of the average potential outcome in the target population is defined as
\begin{equation*}
    \begin{split}
    \widehat{\psi}(a) &= \dfrac{1}{m} \sum\limits_{j = 1}^{m} \Bigg\{ \dfrac{I(S_j = 1, A_j = a)}{\widehat p(X_j, \boldsymbol{W}_j) \widehat e_a(X_j, \boldsymbol{W}_j)}  \Big\{ \overbar{Y}_j - \widehat g_a(X_j, \boldsymbol{W}_j) \Big\} + \widehat g_a(X_j, \boldsymbol{W}_j) \Bigg\}.
    \end{split}
\end{equation*}

Assume that, as $m \rightarrow \infty$, $\widehat p(X, \boldsymbol{W}) \overset{p}{\rightarrow}   \Pr[S = 1 | X,\boldsymbol{W} ] $ and $\widehat e_a(X, \boldsymbol{W}) \overset{p}{\rightarrow}   \Pr[A = a | X, \boldsymbol{W}, S = 1] $  (this assumption is reasonable because both $\Pr[S = 1 | X,\boldsymbol{W} ]$ and $\Pr[A = a | X, \boldsymbol{W}, S = 1] $ are under the evaluators' control and models for them can always be correctly specified). Furthermore, assume that $\widehat g_a(X, \boldsymbol{W}) \overset{p}{\rightarrow} g^*_a(X, \boldsymbol{W})$ where $g^*_a(X, \boldsymbol{W})$ is not necessarily equal to $\E[\overbar{Y} | X,\boldsymbol{W} , S = 1, A = a ]$ (i.e., allowing for misspecification of the outcome model). Then, as the number of clusters grows, 
\begin{equation*}
    \begin{split}
    \widehat{\psi}(a) &\overset{p}{\rightarrow} \E \left [  \dfrac{I(S = 1, A = a) \overbar{Y} }{ \Pr[S = 1 | X,\boldsymbol{W} ] \Pr[A = a | X, \boldsymbol{W}, S = 1]}  \right ] \\
        & \quad\quad\quad +  \E \left [  g^*_a(X, \boldsymbol{W})  \left \{ 1 -  \dfrac{I(S = 1, A = a) }{ \Pr[S = 1 | X,\boldsymbol{W} ] \Pr[A = a | X, \boldsymbol{W}, S = 1]}  \right\}  \right ] .
    \end{split}
\end{equation*}
Using an iterated expectation argument \parencite{dahabreh2019relation} and the definition of $\psi(a)$, we have $$ \E \left [  \dfrac{I(S = 1, A = a) \overbar{Y} }{ \Pr[S = 1 | X,\boldsymbol{W} ] \Pr[A = a | X, \boldsymbol{W}, S = 1]}  \right ] = \E \big[ \E[ \overbar{Y} | X,\boldsymbol{W}, S = 1, A = a ]  \big] \equiv \psi(a).$$ Furthermore, using another iterated expectation argument $$  \E \left [  g^*_a(X, \boldsymbol{W})  \left \{ 1 -  \dfrac{I(S = 1, A = a) }{ \Pr[S = 1 | X,\boldsymbol{W} ] \Pr[A = a | X, \boldsymbol{W}, S = 1]}  \right\}  \right ] = 0 , $$ for all $ g^*_a(X, \boldsymbol{W}) $. 

Thus, we can conclude that, as $m \rightarrow \infty$, $ \widehat{\psi}(a) \overset{p}{\rightarrow}  \psi(a)$, whether $ g^*_a(X, \boldsymbol{W}) $ equals $\E[\overbar{Y} | X,\boldsymbol{W} , S = 1, A = a ]$ or not.

\subsection{Expectation of the average potential outcome in the non-randomized subset of the target population}

The argument is similar to the one for the entire population, given above. Recall that the estimator of the expectation of the average potential outcome  in the target population is defined as
\begin{equation*}
    \begin{split}
    \widehat{\phi}(a) &= \left\{\sum\limits_{j = 1}^{m} I(S_j = 0)\right\}^{-1} \sum\limits_{j = 1}^{m}  \dfrac{I(S_j = 1, A_j = a) \big\{ 1- \widehat p(X_j, \boldsymbol{W}_j) \big\} }{\widehat p(X_j, \boldsymbol{W}_j) \widehat e_a(X_j, \boldsymbol{W}_j)} \Big\{ \overbar{Y}_j -  \widehat g_a(X_j, \boldsymbol{W}_j) \Big\}  \\
    &\quad\quad\quad + \left\{\sum\limits_{j = 1}^{m} I(S_j = 0)\right\}^{-1} \sum\limits_{j = 1}^{m}  \big\{ 1-\widehat p(X_j, \boldsymbol{W}_j) \big\} \widehat g_a(X_j, \boldsymbol{W}_j),
    \end{split}
\end{equation*}

Assume again that, as $m \rightarrow \infty$, $\widehat p(X, \boldsymbol{W}) \overset{p}{\rightarrow}   \Pr[S = 1 | X,\boldsymbol{W} ] $ and $\widehat e_a(X, \boldsymbol{W}) \overset{p}{\rightarrow}  \Pr[A = a | X, \boldsymbol{W}, S = 1] $ (an assumption supported by study design). Furthermore, assume that $\widehat g_a(X, \boldsymbol{W}) \overset{p}{\rightarrow} g^*_a(X, \boldsymbol{W})$ where $g^*_a(X, \boldsymbol{W})$ is not necessarily equal to $\E[\overbar{Y} | X,\boldsymbol{W} , S = 1, A = a ]$ (i.e., allowing for misspecification of the outcome model). Then, as the number of clusters grows, 
\begin{equation*}
    \begin{split}
    \widehat{\phi}(a) &\overset{p}{\rightarrow}  \dfrac{1}{\Pr[S = 0]} \E \left [  \dfrac{I(S = 1, A = a) \overbar{Y} }{ \Pr[S = 1 | X,\boldsymbol{W} ] \Pr[A = a | X, \boldsymbol{W}, S = 1]}  \right ] \\
        & \quad +  \dfrac{1}{\Pr[S = 0]} \E \left [  g^*_a(X, \boldsymbol{W})  \left \{ \Pr[S = 0 | X,\boldsymbol{W} ] -  \dfrac{I(S = 1, A = a) \Pr[S = 0 | X,\boldsymbol{W} ] }{ \Pr[S = 1 | X,\boldsymbol{W} ] \Pr[A = a | X, \boldsymbol{W}, S = 1]}  \right\}  \right ] .
    \end{split}
\end{equation*}
Using an iterated expectation argument \parencite{dahabreh2019relation} and the definition of $\phi(a)$, we have $$ \E \left [  \dfrac{I(S = 1, A = a) \Pr[S = 0 | X,\boldsymbol{W} ] \overbar{Y} }{ \Pr[S = 1 | X,\boldsymbol{W} ] \Pr[A = a | X, \boldsymbol{W}, S = 1]}  \right ] = \E \big[ \E[ \overbar{Y} | X,\boldsymbol{W}, S = 1, A = a ]  \big| S = 0  \big] \equiv \phi(a).$$ Furthermore, using another iterated expectation argument we have $$ \E \left [  g^*_a(X, \boldsymbol{W})  \left \{ \Pr[S = 0 | X,\boldsymbol{W} ] -  \dfrac{I(S = 1, A = a) \Pr[S = 0 | X,\boldsymbol{W} ] }{ \Pr[S = 1 | X,\boldsymbol{W} ] \Pr[A = a | X, \boldsymbol{W}, S = 1]}  \right\}  \right ] = 0 , $$ for all $ g^*_a(X, \boldsymbol{W}) $.

Thus, we can conclude that, as $m \rightarrow \infty$, $ \widehat{\phi}(a) \overset{p}{\rightarrow}  \phi(a)$, whether $ g^*_a(X, \boldsymbol{W}) $ equals $\E[\overbar{Y} | X,\boldsymbol{W} , S = 1, A = a ]$ or not.

\clearpage
\section{Comparison of asymptotic variances when using the known sampling and treatment probabilities}\label{appendix:avar_comp}

\subsection{Expectation of the average potential outcome in the target population}\label{appendix:avar_comp_everyone}

We shall compare the asymptotic variance of the augmented inverse probability weighting estimator $\widehat{\psi}(a)$ (and a version of it using the true sampling probability and the true probability of treatment in the trial) versus the asymptotic variance of a non-augmented inverse probability weighting estimator that uses the true sampling probability and the true probability of treatment in the trial. Here, we assume that $\E[\overbar{Y} |X, \boldsymbol{W} , A = a ]$ is consistently estimated by $\widehat g_a(X,  \boldsymbol{W})$. 

When the sampling $\Pr[S = 1 |  X, \boldsymbol{W} ] = p(X, \boldsymbol{W})$ and treatment probabilities $\Pr[A = a |  X, \boldsymbol{W}, S = 1 ] = e_a( X, \boldsymbol{W}) $ are known, we might consider the augmented inverse probability weighting estimator, 
\begin{equation}
    \begin{split}
    \widecheck{\psi}(a) &= \dfrac{1}{m} \sum\limits_{j = 1}^{m} \Bigg\{ \dfrac{I(S_j = 1, A_j = a)}{p(X_j, \boldsymbol{W}_j) e_a( X_j, \boldsymbol{W}_j) }  \Big\{ \overbar{Y}_j - \widehat g_a(X_j, \boldsymbol{W}_j) \Big\} + \widehat g_a(X_j, \boldsymbol{W}_j) \Bigg\},
    \end{split}
\end{equation}
and the non-augmented inverse probability weighting estimator
\begin{equation}
    \begin{split}
    \widecheck{\psi}_{\text{\tiny w}}(a) &= \dfrac{1}{m} \sum\limits_{j = 1}^{m} \dfrac{I(S_j = 1, A_j = a) \overbar{Y}_j}{p(X_j, \boldsymbol{W}_j) e_a( X_j, \boldsymbol{W}_j)}.
    \end{split}
\end{equation}

Note that the above estimators are obtained from $\widehat{\psi}(a)$ and $\widehat{\psi}_{\text{\tiny w}}(a)$, respectively, by substituting the true (known-by-design) probabilities for the estimated ones. 
The variance of the asymptotic distribution of $\sqrt{m} \big ( \widecheck \psi(a)  - \psi(a) \big)$, which is the same as the variance of the asymptotic distribution of $\sqrt{m} \big ( \widehat{\psi}(a)  - \psi(a) \big)$ \parencite{hahn1998role}, is
\begin{equation*}
    \begin{split}
   V_{\widecheck \psi(a)} &= \E_{p_0} \left[   \Big( \mathit{\Psi}^1_{p_0}(a) \Big)^2 \right] \\
    &= \E_{p_0} \left[ \dfrac{\mbox{Var}_{p_0} [\overbar{Y} |  X, \boldsymbol{W} , S = 1, A = a ] }{\Pr_{p_0}[S = 1 |  X, \boldsymbol{W} ] \Pr_{p_0}[A = a |  X, \boldsymbol{W} , S = 1] } + \big( \E_{p_0}[\overbar{Y} |X, \boldsymbol{W}, S = 1, A = a ] \big)^{2} - \big( \psi_{p_0}(a) \big)^2 \right].
    \end{split}
\end{equation*}
The variance of the asymptotic distribution of $ \sqrt{m} \big ( \widecheck \psi_{\text{\tiny w}}(a) - \psi(a) \big)$, using the results from page 22 of \textcite{tsiatis2007}, is 
\begin{equation*}
    \begin{split}
    V_{\widecheck \psi_{\text{\tiny w}}(a)} &= \E_{p_0} \left[   \left( \dfrac{I(S = 1, A = a)  \overbar{Y} }{\Pr_{p_0}[S = 1 |  X, \boldsymbol{W} ] \Pr_{p_0}[A = a |  X, \boldsymbol{W} , S = 1] }  - \psi_{p_0}(a) \right)^2 \right] \\
    &= \E_{p_0} \left[ \dfrac{\E_{p_0} [\overbar{Y}^2 |  X, \boldsymbol{W} , S = 1, A = a ] }{\Pr_{p_0}[S = 1 |  X, \boldsymbol{W} ] \Pr_{p_0}[A = a |  X, \boldsymbol{W} , S = 1] } - \big( \psi_{p_0}(a) \big)^2 \right],
    \end{split}
\end{equation*}
where the second equality follows from an iterated expectation argument after expanding the square. 

Taking the difference between $V_{\widecheck \psi(a)}$ and $V_{\widecheck \psi_{\text{\tiny w}}(a)}$, and using the fact that $$\mbox{Var}_{p_0} [\overbar{Y} |  X, \boldsymbol{W} , S = 1, A = a ]= \E_{p_0} [\overbar{Y}^2 |  X, \boldsymbol{W} , S = 1, A = a ] - \big( \E_{p_0} [\overbar{Y} |  X, \boldsymbol{W} , S = 1, A = a ] \big)^2,$$
we find that 
\begin{equation*}
    V_{\widecheck \psi(a)}  - V_{\widecheck \psi_{\text{\tiny w}}(a)} = \E \left[  \big ( \E_{p_0} [\overbar{Y} |  X, \boldsymbol{W} , S = 1, A = a ] \big)^2 \left\{ 1 -  \dfrac{1}{\Pr_{p_0}[S = 1 |  X, \boldsymbol{W} ] \Pr_{p_0}[A = a |  X, \boldsymbol{W} , S = 1] } \right\} \right ] \leq 0.
\end{equation*} 
Thus, we conclude that $ V_{\widecheck \psi_{\text{\tiny w}}(a)} \geq  V_{\widecheck \psi(a)}$.

This result suggests that, when the sampling probability and the probability of treatment in the trial are under the control of the evaluators and both the augmented inverse probability weighting estimator and the non-augmented inverse probability weighting estimator are consistent, the augmented estimator (whether using the estimated or true sampling and treatment probabilities) will have asymptotic variance no larger than that of the non-augmented weighting estimator using the true sampling and treatment probabilities, provided the model for $\E [\overbar{Y} |  X, \boldsymbol{W} , S = 1, A = a ]$ is correctly specified. 

Of course, it could be argued that the practical usefulness of this result is limited because the model for the outcome would never be correctly specified in practical applications. Nevertheless, if the model for the outcome can be reasonably approximated (e.g., using machine learning or other data adaptive methods) then we would expect the asymptotic variance of the augmented inverse probability weighting estimator to be smaller than that of the non-augmented weighting estimator. Furthermore, any misspecification of the outcome model is unlikely to prove severely detrimental in large samples because of the robustness property of the augmented inverse probability weighting estimator.

\subsection{Expectation of the average potential outcome in the non-randomized subset of the target population}

\paragraph{Comparing the non-augmented inverse odds weighting estimator vs augmented inverse odds weighting estimator:} A similar analytical result as the one presented in Section \ref{appendix:avar_comp_everyone} is not available for estimators of $\phi(a)$. Nevertheless, when the sampling probabilities $\Pr[S = 1 |  X = x, \boldsymbol{W} = \boldsymbol{w} ]$ are small for all covariate patterns $\boldsymbol{x}, \boldsymbol{w}$ (as will often be the case), we should expect the behavior of (augmented) inverse odds weighting estimators to be similar to that of (augmented) inverse probability weighting estimators.

For completeness, we provide the asymptotic variance of the augmented inverse odds weighting estimator versus that of the non-augmented inverse odds weighting estimator when the sampling and treatment probabilities are known, as would be the case in a designed study where clusters are sampled with known sampling probability and the treatment assignment is under the control of the investigators. Here, we assume that $\E[\overbar{Y} |X, \boldsymbol{W} , S = 1, A = a ]$ can be estimated consistently by $\widehat g_a(X,  \boldsymbol{W})$. 

When the sampling $\Pr[S = 1 |  X, \boldsymbol{W} ] = p(X, \boldsymbol{W})$ and treatment probabilities $\Pr[A = a |  X, \boldsymbol{W}, S = 1 ] = e_a( X, \boldsymbol{W}) $ are known, we might consider the following two estimators: an augmented inverse odds weighting estimator for $\phi(a)$, 
\begin{equation}\label{appendix_eq:AIOW_est_S0}
    \begin{split}
    \widecheck{\phi}(a) &= \left\{\sum\limits_{j = 1}^{m} I(S_j = 0)\right\}^{-1} \sum\limits_{j = 1}^{m}  w_a(X_j, \boldsymbol{W}_j, S_j, A_j ) \Big\{ \overbar{Y}_j -  \widehat g_a(X_j, \boldsymbol{W}_j) \Big\}  \\
    &\quad\quad\quad + \left\{\sum\limits_{j = 1}^{m} I(S_j = 0)\right\}^{-1} \sum\limits_{j = 1}^{m}  \big\{ 1- p(X_j, \boldsymbol{W}_j) \big\} \widehat g_a(X_j, \boldsymbol{W}_j),
    \end{split}
\end{equation}
with $$  w_a(X, \boldsymbol{W}, S, A ) =  \dfrac{I(S = 1, A = a) \big\{ 1- p(X, \boldsymbol{W}) \big\} }{p(X, \boldsymbol{W}) e_a( X, \boldsymbol{W})},$$ and a non-augmented inverse odds weighting estimator for $\phi(a)$,
\begin{equation}\label{appendix_eq:IOW_est_S0}
    \begin{split}
    \widecheck{\phi}_{\text{\tiny w}}(a) &= \left\{\sum\limits_{j = 1}^{m} I(S_j = 0)\right\}^{-1} \sum\limits_{j = 1}^{m}  w_a(X_j, \boldsymbol{W}_j, S_j, A_j )  \overbar{Y}_j. 
    \end{split}
\end{equation}
Note that the above augmented inverse odds weighting estimator and non-augmented inverse odds weighting estimator are obtained from $\widehat{\phi}(a)$ and $\widehat{\phi}_{\text{\tiny w}}(a)$, respectively, by substituting the true probabilities for the estimated ones. 

The asymptotic variance of the distribution of $\sqrt{m} \big ( \widecheck{\phi}(a) - \phi(a) \big )$ is 
\begin{equation*}\label{eq:var_bound_phi}
  \begin{split}
   V_{\widecheck \phi(a)} &=
  \E\left[ \big( \widetilde{\mathit{\Phi}}^1_{p_0}(a) \big)^2 \right]  \\
  &= \quad \frac{1}{\pi_{p_0}^2}  \Bigg\{ \E \left[ \dfrac{ \big(1 - \Pr_{p_0}[ S = 1 | X, \boldsymbol{W}]\big)^2 \mbox{Var}_{p_0} [\overbar{Y} |  X, \boldsymbol{W} , S = 1, A = a ]}{\Pr_{p_0}[S = 1 |  X, \boldsymbol{W} ] \Pr_{p_0}[A = a |  X, \boldsymbol{W} , S = 1]  }  \right] \\
  &\quad\quad\quad\quad\quad\quad\quad\quad\quad\quad\quad\quad + \E\Big[ (1-{\Pr}_{p_0}[S = 1|X, \boldsymbol{W}])^2 \big\{ \E_{p_0}[\overbar{Y} |  X, \boldsymbol{W} , S = 1, A = a] - \phi_{p_0}(a) \big\}^2 \Big] \Bigg \}.
  \end{split}
\end{equation*}

The asymptotic variance of $\sqrt{m} \big ( \widecheck{\phi}_{\text{\tiny w}}(a) - \phi(a) \big )$, applying the results from page 22 of \textcite{tsiatis2007}, is 
\begin{equation*}
    \begin{split}
    V_{\widecheck \phi_{\text{\tiny w}}(a)} &= \frac{1}{\pi_{p_0}^2}  \E_{p_0} \left[   \left( \dfrac{I(S = 1, A = a)  \overbar{Y} (1 - \Pr_{p_0}[ S = 1 | X,\boldsymbol{W}])  }{\Pr_{p_0}[S = 1 |  X, \boldsymbol{W} ] \Pr_{p_0}[A = a |  X, \boldsymbol{W} , S = 1] }  - I(S = 0) \phi_{p_0}(a) \right)^2 \right]. \\
    \end{split}
\end{equation*}

\paragraph{Comparing the augmented inverse odds weighting estimators:} For completeness, we shall compare the asymptotic variance of the augmented inverse odds weighting estimator $\widecheck{\phi}(a)$ using the true sampling probability and the true probability of treatment in the trial, from equation \eqref{appendix_eq:AIOW_est_S0} (corresponding to the influence function,  $\widetilde{\mathit{\Phi}}^1_{p_0}(a)$, in Appendix B) versus the asymptotic variance of an augmented inverse odds weighting estimator that uses the true sampling probability and the true probability of treatment in the trial (corresponding to the influence function, $\mathit{\Phi}^1_{p_0}(a)$ in the technical report and also given in Appendix B):

\begin{equation}\label{appendix_eq:AIOW_Stat_Med_technical_unpublished}
    \begin{split}
    \bar{\phi}(a) &= \left\{\sum\limits_{j = 1}^{m} I(S_j = 0)\right\}^{-1} \sum\limits_{j = 1}^{m}  w_a(X_j, \boldsymbol{W}_j, S_j, A_j ) \Big\{ \overbar{Y}_j -  \widehat g_a(X_j, \boldsymbol{W}_j) \Big\}  \\
    &\quad\quad\quad + \left\{\sum\limits_{j = 1}^{m} I(S_j = 0)\right\}^{-1} \sum\limits_{j = 1}^{m}  (1- S_j) \widehat g_a(X_j, \boldsymbol{W}_j),
    \end{split}
\end{equation}
with $$  w_a(X, \boldsymbol{W}, S, A ) =  \dfrac{I(S = 1, A = a) \big\{ 1- \Pr[S = 1 |  X, \boldsymbol{W} ] \big\} }{\Pr[S = 1 |  X, \boldsymbol{W} ] \Pr[A = a |  X, \boldsymbol{W}, S = 1 ]}.$$

Here, we assume that $\E[\overbar{Y} |X, \boldsymbol{W} , S = 1, A = a ]$ can be consistently estimated by $\widehat g_a(X,  \boldsymbol{W})$. The asymptotic variance of the distribution of $\sqrt{m} \big ( \widecheck{\phi}(a) - \phi(a) \big )$ \parencite{hahn1998role} is 
\begin{equation*}\label{eq:var_bound_phi}
  \begin{split}
   V_{\widecheck \phi(a)} &=
  \E\left[ \big( \widetilde{\mathit{\Phi}}^1_{p_0}(a) \big)^2 \right]  \\
  &= \quad \frac{1}{\pi_{p_0}^2}  \Bigg\{ \E \left[ \dfrac{ \big(1 - \Pr_{p_0}[ S = 1 | X, \boldsymbol{W}]\big)^2 \mbox{Var}_{p_0} [\overbar{Y} |  X, \boldsymbol{W} , S = 1, A = a ]}{\Pr_{p_0}[S = 1 |  X, \boldsymbol{W} ] \Pr_{p_0}[A = a |  X, \boldsymbol{W} , S = 1]  }  \right] \\
  &\quad\quad\quad\quad\quad\quad\quad\quad\quad +  \E \Big[ (1- {\Pr}_{p_0}[S = 1 |  X, \boldsymbol{W} ])^2 \big\{ \E_{p_0}[\overbar{Y} |  X, \boldsymbol{W} , S = 1, A = a] - \phi_{p_0}(a) \big\}^2  \Big]  \Bigg \}.
  \end{split}
\end{equation*}

Taking the difference between $V_{\widecheck \phi(a)}$ and $V_{\bar \phi(a)}$, 
\begin{equation*}
 \begin{split}
     V_{\widecheck \phi(a)} - V_{\bar \phi(a)}  &= \E \left[ (1-{\Pr}_{p_0}[S = 1|X, \boldsymbol{W}])^2 \big\{ \E_{p_0}[\overbar{Y} |  X, \boldsymbol{W} , S = 1, A = a] - \phi_{p_0}(a) \big\}^2  \right ] \\
     &\quad\quad\quad - \E \left[ (1-S) \big\{ \E_{p_0}[\overbar{Y} |  X, \boldsymbol{W} , S = 1, A = a]  - \phi_{p_0}(a) \big\}^2  \right ] \\
    &= \E \left[ (1-{\Pr}_{p_0}[S = 1|X, \boldsymbol{W}])^2 \big\{ \E_{p_0}[\overbar{Y} |  X, \boldsymbol{W} , S = 1, A = a] - \phi_{p_0}(a) \big\}^2  \right ] \\
    &\quad\quad\quad -  \E \left[ (1-{\Pr}_{p_0}[S = 1|X, \boldsymbol{W}]) \big\{ \E_{p_0}[\overbar{Y} |  X, \boldsymbol{W} , S = 1, A = a]  - \phi_{p_0}(a) \big\}^2  \right ] \\
    &= - \E \left[ {\Pr}_{p_0}[S = 1|X, \boldsymbol{W}]  (1-{\Pr}_{p_0}[S = 1|X, \boldsymbol{W}]) \big\{ \E_{p_0}[\overbar{Y} |  X, \boldsymbol{W} , S = 1, A = a]  - \phi_{p_0}(a) \big\}^2  \right ]\\
    &\leq 0 .
     \end{split}
\end{equation*} 
Thus, we conclude that $V_{\widecheck \phi(a)} \leq V_{\bar \phi(a)}$. 

\clearpage
\section{Simulation scenario 1 in the non-randomized subset}\label{appendix_sim_tables_non_randomized}
\setcounter{table}{0}
\renewcommand{\tablename}{Appendix Table}


\renewcommand{\arraystretch}{1.16}
\begin{table}[ht!]
\centering
\caption{Scaled bias of estimators for quantities in the non-randomized subset of the target population.}
\label{table_selected_results_simulation_targetS0}
\small
\begin{tabular}{lcccccc}
\toprule
Estimand & $n$   & Values for probabilities & Trial-only  & IOW     & AIOW1   & AIOW2   \\ \midrule
ATE in $S = 0$       & 50                        & True                & 12.409 & -0.045 & 0.024  & 0.025  \\
     & 50                        & Estimated (simple)  & 12.409 & 0.015  & 0.025  & 0.025  \\
     & 50                        & Estimated (complex) & 12.409 & 0.001  & -0.019 & 0.025  \\
      & 100                       & True                & 12.396 & -0.233 & 0.008  & 0.012  \\
      & 100                       & Estimated (simple)  & 12.396 & -0.049 & 0.009  & 0.013  \\
     & 100                       & Estimated (complex) & 12.396 & 0.047  & -0.005 & 0.016  \\
     & 200                       & True                & 12.807 & 0.130   & 0.002  & 0.002  \\
      & 200                       & Estimated (simple)  & 12.807 & 0.012  & 0.001  & 0.001  \\
      & 200                       & Estimated (complex) & 12.807 & -0.028 & -0.013 & -0.001 \\ \midrule
$\E \big [\overbar{Y}^{a=1} \big | S = 0 \big]$       & 50                        & True                & 6.160   & -0.138 & 0.012  & 0.013  \\
      & 50                        & Estimated (simple)  & 6.160   & -0.024 & 0.013  & 0.014  \\
      & 50                        & Estimated (complex) & 6.160   & 0.041  & -0.005 & 0.011  \\
      & 100                       & True                & 6.185  & -0.199 & -0.002 & 0.004  \\
      & 100                       & Estimated (simple)  & 6.185  & -0.061 & 0.000      & 0.006  \\
     & 100                       & Estimated (complex) & 6.185  & -0.011 & -0.007 & 0.008  \\
      & 200                       & True                & 6.398  & 0.020   & -0.005 & -0.005 \\
      & 200                       & Estimated (simple)  & 6.398  & -0.015 & -0.005 & -0.005 \\
      & 200                       & Estimated (complex) & 6.398  & -0.037 & -0.011 & -0.005 \\ \midrule
$\E \big [\overbar{Y}^{a=0} \big | S = 0 \big]$       & 50                        & True                & -6.249 & -0.093 & -0.012 & -0.012 \\
      & 50                        & Estimated (simple)  & -6.249 & -0.039 & -0.011 & -0.012 \\
      & 50                        & Estimated (complex) & -6.249 & 0.040   & 0.014  & -0.014 \\
      & 100                       & True                & -6.211 & 0.034  & -0.010  & -0.008 \\
      & 100                       & Estimated (simple)  & -6.211 & -0.011 & -0.009 & -0.007 \\
      & 100                       & Estimated (complex) & -6.211 & -0.058 & -0.001 & -0.008 \\
      & 200                       & True                & -6.409 & -0.111 & -0.006 & -0.007 \\
      & 200                       & Estimated (simple)  & -6.409 & -0.027 & -0.006 & -0.006 \\
      & 200                       & Estimated (complex) & -6.409 & -0.009 & 0.002  & -0.004 \\ \bottomrule
\end{tabular}
\caption*{Results are scaled by $\sqrt{m}$ (i.e., multiplied by $\sqrt{5000} \approx 70.7$). ATE in $S=0$ is defined as $\E \big [\overbar{Y}^{a=1} \big | S = 0 \big] - \E \big [\overbar{Y}^{a=0} \big | S = 0 \big] $; $n$ = number of clusters in the trial; Values for probabilities = how the treatment and sampling probabilities are obtained for the estimators (True = use the known-by-design sampling probabilities and probabilities of treatment in the trial; Estimated (simple) = estimate the sampling and treatment probabilities conditional on $X_j$ only (the variable used to determine the sampling probabilities); Estimated (complex) = estimate the sampling and treatment probabilities conditional on $X_j$ and the cluster-level averages of $\boldsymbol{W}_{1,j}$ and $\boldsymbol{W}_{2,j}$); Trial-only = average the individual-level outcomes in each cluster and then take the average of these averages over the clusters participating in the trial; IOW = non-augmented inverse odds weighting estimator; AIOW1 = augmented inverse odds weighting estimator, with the outcome model fit only at the cluster-level; AIOW2 = augmented inverse odds weighting estimator, with the outcome model fit at the individual-level.}
\end{table}
\renewcommand{\arraystretch}{1}

\renewcommand{\arraystretch}{1.16}
\begin{table}[ht!]
\centering
\caption{Scaled standard deviation of estimators for quantities in the non-randomized subset of the target population.}
\label{table_selected_results_simulation_targetS0_SD}
\begin{tabular}{lccccc}
\toprule
Estimand & $n$   & Values for probabilities & IOW & AIOW1 & AIOW2 \\ \midrule
ATE in $S = 0$       & 50                        & True                & 14.198 & 1.459    & 1.39     \\
     & 50                        & Estimated (simple)  & 4.869  & 1.460     & 1.392    \\
      & 50                        & Estimated (complex) & 5.162  & 1.492    & 1.436    \\
     & 100                       & True                & 10.237 & 0.997    & 0.952    \\
     & 100                       & Estimated (simple)  & 3.429  & 1.000        & 0.954    \\
      & 100                       & Estimated (complex) & 2.862  & 1.005    & 0.968    \\
      & 200                       & True                & 7.335  & 0.749    & 0.721    \\
      & 200                       & Estimated (simple)  & 2.351  & 0.751    & 0.722    \\
     & 200                       & Estimated (complex) & 1.706  & 0.749    & 0.729    \\ \midrule
$\E \big [\overbar{Y}^{a=1} \big | S = 0 \big]$        & 50                        & True                & 9.956  & 1.056    & 0.994    \\
     & 50                        & Estimated (simple)  & 3.480   & 1.055    & 0.994    \\
      & 50                        & Estimated (complex) & 3.814  & 1.078    & 1.025    \\
      & 100                       & True                & 7.173  & 0.679    & 0.645    \\
      & 100                       & Estimated (simple)  & 2.448  & 0.678    & 0.644    \\
     & 100                       & Estimated (complex) & 2.119  & 0.683    & 0.655    \\
      & 200                       & True                & 5.238  & 0.519    & 0.495    \\
      & 200                       & Estimated (simple)  & 1.720   & 0.512    & 0.488    \\
     & 200                       & Estimated (complex) & 1.279  & 0.507    & 0.491    \\ \midrule 
$\E \big [\overbar{Y}^{a=0} \big | S = 0 \big]$        & 50                        & True                & 10.159 & 0.989    & 0.942    \\
     & 50                        & Estimated (simple)  & 3.460   & 0.988    & 0.943    \\
      & 50                        & Estimated (complex) & 3.833  & 0.997    & 0.965    \\
      & 100                       & True                & 7.179  & 0.703    & 0.660     \\
      & 100                       & Estimated (simple)  & 2.503  & 0.700      & 0.656    \\
      & 100                       & Estimated (complex) & 2.097  & 0.696    & 0.664    \\
      & 200                       & True                & 5.169  & 0.502    & 0.482    \\
     & 200                       & Estimated (simple)  & 1.682  & 0.495    & 0.475    \\
      & 200                       & Estimated (complex) & 1.282  & 0.491    & 0.477    \\ \bottomrule
\end{tabular}
\caption*{Results are scaled by $\sqrt{m}$ (i.e., multiplied by $\sqrt{5000} \approx 70.7$). ATE in $S = 0$ is defined as $\E \big [\overbar{Y}^{a=1} \big | S = 0 \big] - \E \big [\overbar{Y}^{a=0} \big | S = 0 \big] $; $n$ = number of clusters in the trial; Values for probabilities = how the treatment and sampling probabilities are obtained for the estimators (True = use the known-by-design sampling probabilities and probabilities of treatment in the trial; Estimated (simple) = estimate the sampling and treatment probabilities conditional on $X_j$ only (the variable used to determine the sampling probabilities); Estimated (complex) = estimate the sampling and treatment probabilities conditional on $X_j$ and the cluster-level averages of $\boldsymbol{W}_{1,j}$ and $\boldsymbol{W}_{2,j}$); IOW = non-augmented inverse odds weighting estimator; AIOW1 = augmented inverse odds weighting estimator, with the outcome model fit only at the cluster-level; AIOW2 = augmented inverse odds weighting estimator, with the outcome model fit at the individual-level.}
\end{table}
\renewcommand{\arraystretch}{1}

\begin{table}[ht!]
\centering
\caption{Coverage and scaled average of standard errors for the augmented weighting estimators for quantities in the non-randomized subset of the target population.}
\label{table_selected_results_simulation_targetS0_coverage}
\resizebox{\textwidth}{!}{

\renewcommand{\arraystretch}{1.16}
\begin{tabular}{lcccccccccc@{}}
\toprule
\multicolumn{1}{l}{}               & \multicolumn{1}{c}{}    & \multicolumn{1}{c}{}     & \multicolumn{4}{c}{AIOW1}                                                                         & \multicolumn{4}{c}{AIOW2}   \\ [0.1cm]
\multicolumn{1}{l}{Estimand}       & \multicolumn{1}{c}{$n$} & \multicolumn{1}{c}{ Values for probabilities} & \multicolumn{2}{c}{ASE}                          & \multicolumn{2}{c}{Coverage}                    & \multicolumn{2}{c}{ASE}                          & \multicolumn{2}{c}{Coverage}                    \\ [0.1cm]
\multicolumn{1}{l}{}               & \multicolumn{1}{c}{}    & \multicolumn{1}{c}{}     & \multicolumn{1}{c}{IC} & \multicolumn{1}{c}{BS} & \multicolumn{1}{c}{IC} & \multicolumn{1}{c}{BS} & \multicolumn{1}{c}{IC} & \multicolumn{1}{c}{BS} & \multicolumn{1}{c}{IC} & \multicolumn{1}{c}{BS} \\ \midrule
ATE in $S = 0$       & 50  & True                & 1.210          & 1.784         & 0.877          & 0.969         & 1.148         & 1.524         & 0.885         & 0.959          \\
     & 50  & Estimated (simple)  & 1.251         & 1.785         & 0.906          & 0.969          & 1.181         & 1.525         & 0.900            & 0.960        \\
      & 50  & Estimated (complex) & 1.297         & 1.920          & 0.903          & 0.963        & 1.215         & 1.634         & 0.899         & 0.959        \\
      & 100 & True                & 0.934         & 1.029         & 0.924         & 0.950           & 0.891         & 0.971         & 0.9295         & 0.952         \\
      & 100 & Estimated (simple)  & 0.951         & 1.030          & 0.935        & 0.950           & 0.906         & 0.972         & 0.937         & 0.952         \\
      & 100 & Estimated (complex) & 0.971         & 1.031         & 0.936         & 0.953         & 0.922         & 0.985         & 0.938          & 0.954          \\
      & 200 & True                & 0.722         & 0.751         & 0.932          & 0.943         & 0.694         & 0.717         & 0.933          & 0.946          \\
      & 200 & Estimated (simple)  & 0.729         & 0.753         & 0.939        & 0.946          & 0.699         & 0.719         & 0.940           & 0.944          \\
      & 200 & Estimated (complex) & 0.736         & 0.747         & 0.940           & 0.941          & 0.706         & 0.725         & 0.942          & 0.949          \\ \midrule 
$\E \big [\overbar{Y}^{a=1} \big | S = 0 \big]$       & 50  & True                & 0.826         & 1.194         & 0.85           & 0.9505         & 0.780          & 1.039         & 0.856         & 0.9445         \\
      & 50  & Estimated (simple)  & 0.856         & 1.194         & 0.8785         & 0.9495         & 0.807         & 1.038         & 0.8805         & 0.9435         \\
     & 50  & Estimated (complex) & 0.885         & 1.268         & 0.879          & 0.958          & 0.828         & 1.101         & 0.876          & 0.945          \\
      & 100 & True                & 0.632         & 0.706         & 0.9135         & 0.952          & 0.600           & 0.664         & 0.9125         & 0.9485         \\
     & 100 & Estimated (simple)  & 0.647         & 0.703         & 0.935        & 0.951          & 0.613         & 0.660          & 0.9315         & 0.948          \\
      & 100 & Estimated (complex) & 0.661         & 0.705         & 0.932          & 0.946          & 0.625         & 0.669         & 0.9275         & 0.947          \\
      & 200 & True                & 0.480         & 0.509         & 0.921         & 0.940           & 0.458         & 0.485         & 0.923         & 0.944         \\
      & 200 & Estimated (simple)  & 0.484         & 0.501         & 0.9295         & 0.938         & 0.462         & 0.476         & 0.931         & 0.944         \\
      & 200 & Estimated (complex) & 0.489         & 0.498         & 0.936          & 0.937          & 0.466         & 0.479         & 0.935          & 0.942         \\ \midrule
$\E \big [\overbar{Y}^{a=0} \big | S = 0 \big]$       & 50  & True                & 0.823         & 1.191         & 0.867         & 0.950           & 0.777         & 1.039         & 0.858         & 0.952         \\
      & 50  & Estimated (simple)  & 0.852         & 1.191         & 0.890         & 0.950         & 0.803         & 1.038         & 0.889          & 0.950           \\
     & 50  & Estimated (complex) & 0.881         & 1.282         & 0.8935         & 0.9555         & 0.823         & 1.111         & 0.888          & 0.953          \\
      & 100 & True                & 0.634         & 0.704         & 0.902         & 0.941          & 0.603         & 0.663         & 0.909         & 0.941         \\
    & 100 & Estimated (simple)  & 0.644         & 0.701         & 0.920         & 0.943        & 0.611         & 0.660          & 0.923         & 0.941          \\
     & 100 & Estimated (complex) & 0.657         & 0.701         & 0.927          & 0.941          & 0.622         & 0.667         & 0.926          & 0.942          \\
      & 200 & True                & 0.479         & 0.510          & 0.9305         & 0.947          & 0.457         & 0.485         & 0.921         & 0.946         \\
      & 200 & Estimated (simple)  & 0.485         & 0.502         & 0.933          & 0.943          & 0.463         & 0.476         & 0.936         & 0.943          \\
     & 200 & Estimated (complex) & 0.491         & 0.498         & 0.935          & 0.941          & 0.467         & 0.480          & 0.936         & 0.938          \\ \bottomrule
\end{tabular}

}
\caption*{ASE = average (over the simulations) of standard errors, scaled by $\sqrt{m}$ (i.e., multiplied by $\sqrt{5000} \approx 70.7$); IC = influence-curve based; BS = bootstrap; Coverage = coverage, using 95\% normal confidence intervals. ATE in $S = 0$ is defined as $\E \big [\overbar{Y}^{a=1}  \big | S = 0 \big] - \E \big [\overbar{Y}^{a=0} \big | S = 0 \big] $; $n$ = number of clusters in the trial; Values for probabilities = how the treatment and sampling probabilities are obtained for the estimators (True = use the known-by-design sampling probabilities and probabilities of treatment in the trial; Estimated (simple) = estimate the sampling and treatment probabilities conditional on $X_j$ only (the variable used to determine the sampling probabilities); Estimated (complex) = estimate the sampling and treatment probabilities conditional on $X_j$ and the cluster-level averages of $\boldsymbol{W}_{1,j}$ and $\boldsymbol{W}_{2,j}$); AIOW1 = augmented inverse odds weighting estimator, with the outcome model fit only at the cluster-level; AIOW2 = augmented inverse odds weighting estimator, with the outcome model fit at the individual-level.}
\end{table}
\renewcommand{\arraystretch}{1}

\clearpage
\section{Details for additional simulation scenarios}\label{appendix_addtional_sim_details}

All simulation scenarios are modifications of Scenario 1, as summarized in Table 1 of the main text. Here, we provide additional details for scenarios that required modifications to the generative model for the outcome or to the specification models used in analyzing the simulated data (see Appendix Table \ref{AppendixTable:Scenarios_Model} for a summary).

\subsection{Generative models for the outcome}

In Scenario 2, we examined the impact of having no interference by generating outcomes under a logistic model with the following linear predictor:
$$L_{j,i}= (2 A_j - 1) X_{j} + 0.5(2 A_j - 1) W_{1, j, i} + 0.5(2A_j-1) W_{2, j, i}.$$ 

In Scenario 3, we considered weaker treatment effects and weaker heterogeneity, by generating outcomes under a logistic model with the following linear predictor:
$$L_{j,i}=0.5 \Big( (2 A_j - 1) X_{j} + 0.5(2 A_j - 1) W_{1, j, i} + 0.5(2A_j-1) W_{2, j, i} +0.5(2A_j-1)\overbar{W}_{1,j}+0.5(2A_j-1)\overbar{W}_{2,j} \Big).$$ 


In Scenario 4, we generated data under the sharp causal null hypothesis (no treatment effect and no heterogeneity), by generating outcomes under a logistic model with the following linear predictor:
$$L_{j,i}= -X_{j} - 0.5 W_{1, j, i} -0.5 W_{2, j, i}-0.5\overbar{W}_{1,j} -0.5\overbar{W}_{2,j} .$$

In Scenario 5, we used the same generative model for the outcome as in Scenario 1.

In Scenario 6, we considered nonlinearity (on the logit scale) by generating outcomes under a logistic model with the following linear predictor: $$L_{j,i}= (2 A_j - 1) X_{j} + 0.5(2 A_j - 1) W_{1, j, i}^2 + 0.5(2A_j-1) W_{2, j, i}^2 +0.5(2A_j-1)\overbar{W}_{1,j}^2+0.5(2A_j-1)\overbar{W}_{2,j}^2.$$

\vspace{1in}
\begin{table}[ht!]
\centering
\caption{Summary of outcome model data generation mechanisms and outcome model specifications used in different scenarios.}
\label{AppendixTable:Scenarios_Model}
\renewcommand{\arraystretch}{1.75}
\begin{tabular}{@{}cccc@{}}
\toprule
Scenario & True DGM                  & Individual-level outcome model & Complex participation model \\ \midrule
1        & $W_{1,j}, W_{2,j}, X_j, \overbar{W}_{1,j}, \overbar{W}_{2,j}$         & $W_{1,j}, W_{2,j}, X_j, \overbar{W}_{1,j}, \overbar{W}_{2,j}$                           & $X_j, \overbar{W}_{1,j}, \overbar{W}_{2,j}$                   \\
2        &$W_{1,j}, W_{2,j}, X_j$             & $W_{1,j}, W_{2,j}, X_j, \overbar{W}_{1,j}, \overbar{W}_{2,j}$             &  $X_j, \overbar{W}_{1,j}, \overbar{W}_{2,j}$                     \\
3        & $W_{1,j}, W_{2,j}, X_j, \overbar{W}_{1,j}, \overbar{W}_{2,j}$          & $W_{1,j}, W_{2,j}, X_j, \overbar{W}_{1,j}, \overbar{W}_{2,j}$                             &  $X_j, \overbar{W}_{1,j}, \overbar{W}_{2,j}$                     \\
4        & $W_{1,j}, W_{2,j}, X_j, \overbar{W}_{1,j}, \overbar{W}_{2,j}$          & $W_{1,j}, W_{2,j}, X_j, \overbar{W}_{1,j}, \overbar{W}_{2,j}$                             &  $X_j, \overbar{W}_{1,j}, \overbar{W}_{2,j}$                    \\
5        & $W_{1,j}, W_{2,j}, X_j, \overbar{W}_{1,j}, \overbar{W}_{2,j}$          &  $W_{1,j}, X_j, \overbar{W}_{1,j}$                         &  $X_j, \overbar{W}_{1,j}$                         \\
6        & $W^2_{1,j}, W^2_{2,j}, X_j, \overbar{W}^2_{1,j}, \overbar{W}^2_{2,j}$  & $W_{1,j}, W_{2,j}, X_j, \overbar{W}_{1,j}, \overbar{W}_{2,j}$                 &  $X_j, \overbar{W}_{1,j}, \overbar{W}_{2,j}$                 \\ \bottomrule
\end{tabular}
\caption*{ Scenario = scenarios as defined in Table 1 of the main text; True DGM = True data generating mechanism used to generate the individual-level observed outcome; Individual-level outcome model = model specification used to estimate the outcome model at the individual-level, where same indicates the same variables were used in the DGM; Complex participation model = estimated complex participation model (fit only at the cluster-level). The treatment model uses the same specification as the participation model. }
\end{table}

\clearpage
\section{Simulation scenario 2} \label{appendix_scenario2}

\renewcommand{\arraystretch}{1.16}
\begin{table}[ht!]
\centering
\caption{Scaled bias of estimators for quantities in the target population.}
\label{table_selected_results_simulation_targetALL_scenario}
\small
\begin{tabular}{lcccccc}
\toprule
Estimand & $n$   & Values for probabilities & Trial-only  & IPW     & AIPW1   & AIPW2   \\ \midrule
ATE      & 50                        & True                & 13.204 & -0.020 & 0.034  & 0.034  \\
    & 50                        & Estimated (simple)  & 13.204 & 0.037  & 0.034  & 0.034  \\
    & 50                        & Estimated (complex) & 13.204 & 0.028  & 0.018  & 0.031  \\
      & 100                       & True                & 13.083 & -0.104 & 0.003  & 0.001  \\
      & 100                       & Estimated (simple)  & 13.083 & 0.017  & 0.003  & 0.002  \\
      & 100                       & Estimated (complex) & 13.083 & 0.017  & 0.002  & 0.007  \\
     & 200                       & True                & 13.147 & -0.330 & 0.006  & 0.009  \\
    & 200                       & Estimated (simple)  & 13.147 & -0.015 & 0.006  & 0.009  \\
     & 200                       & Estimated (complex) & 13.147 & -0.035 & 0.001  & 0.007  \\ \midrule 
$\E \big [\overbar{Y}^{a=1} \big]$       & 50                        & True                & 6.586  & -0.102 & 0.023  & 0.025  \\
      & 50                        & Estimated (simple)  & 6.586  & 0.006  & 0.023  & 0.024  \\
    & 50                        & Estimated (complex) & 6.586  & 0.036  & 0.018  & 0.022  \\
     & 100                       & True                & 6.574  & 0.050  & -0.008 & -0.012 \\
     & 100                       & Estimated (simple)  & 6.574  & 0.028  & -0.008 & -0.011 \\
     & 100                       & Estimated (complex) & 6.574  & 0.019  & -0.006 & -0.006 \\
     & 200                       & True                & 6.592  & -0.169 & 0.006  & 0.008  \\
     & 200                       & Estimated (simple)  & 6.592  & 0.001  & 0.006  & 0.008  \\
     & 200                       & Estimated (complex) & 6.592  & -0.017 & 0.004  & 0.008  \\ \midrule 
$\E \big [\overbar{Y}^{a=0} \big]$         & 50                        & True                & -6.618 & -0.082 & -0.010 & -0.009 \\
    & 50                        & Estimated (simple)  & -6.618 & -0.032 & -0.010 & -0.009 \\
     & 50                        & Estimated (complex) & -6.618 & 0.009  & 0.000  & -0.009 \\
    & 100                       & True                & -6.509 & 0.154  & -0.011 & -0.013 \\
     & 100                       & Estimated (simple)  & -6.509 & 0.010  & -0.011 & -0.013 \\
      & 100                       & Estimated (complex) & -6.509 & 0.002  & -0.009 & -0.013 \\
      & 200                       & True                & -6.555 & 0.161  & 0.000  & -0.001 \\
     & 200                       & Estimated (simple)  & -6.555 & 0.015  & 0.000  & -0.001 \\
     & 200                       & Estimated (complex) & -6.555 & 0.017  & 0.003  & 0.000  \\ \bottomrule
\end{tabular}
\caption*{Results are scaled by $\sqrt{m}$ (i.e., multiplied by $\sqrt{5000} \approx 70.7$). ATE is defined as $\E \big [\overbar{Y}^{a=1} \big] - \E \big [\overbar{Y}^{a=0} \big] $; $n$ = number of clusters in the trial; Values for probabilities = how the treatment and sampling probabilities are obtained for the estimators (True = use the known-by-design sampling probabilities and probabilities of treatment in the trial; Estimated (simple) = estimate the sampling and treatment probabilities conditional on $X_j$ only (the variable used to determine the sampling probabilities); Estimated (complex) = estimate the sampling and treatment probabilities conditional on $X_j$ and the cluster-level averages of $\boldsymbol{W}_{1,j}$ and $\boldsymbol{W}_{2,j}$); Trial-only = average the individual-level outcomes in each cluster and then take the average of these averages over the clusters participating in the trial; IPW = non-augmented inverse probability weighting estimator; AIPW1 = augmented inverse probability weighting estimator, with the outcome model fit only at the cluster-level; AIPW2 = augmented inverse probability weighting estimator, with the outcome model fit at the individual-level.}
\end{table}
\renewcommand{\arraystretch}{1}

\renewcommand{\arraystretch}{1.16}
\begin{table}[ht!]
\centering
\caption{Scaled standard deviation of estimators for quantities in the target population.}
\label{table_selected_results_simulation_targetALL_SD_appendix_scenario2}
\begin{tabular}{lccccc}
\toprule
Estimand & $n$   & Values for probabilities & IPW & AIPW1 & AIPW2 \\ \midrule
ATE      & 50                        & True                & 13.629 & 1.402    & 1.397    \\
     & 50                        & Estimated (simple)  & 2.813  & 1.402    & 1.398    \\
      & 50                        & Estimated (complex) & 4.098  & 1.446    & 1.440    \\
     & 100                       & True                & 9.524  & 0.956    & 0.942    \\
     & 100                       & Estimated (simple)  & 1.991  & 0.956    & 0.941    \\
     & 100                       & Estimated (complex) & 1.894  & 0.959    & 0.949    \\
     & 200                       & True                & 6.910  & 0.681    & 0.676    \\
     & 200                       & Estimated (simple)  & 1.400  & 0.681    & 0.676    \\
      & 200                       & Estimated (complex) & 1.133  & 0.682    & 0.679    \\ \midrule
$\E \big [\overbar{Y}^{a=1} \big]$      & 50                        & True                & 9.573  & 1.011    & 1.008    \\
     & 50                        & Estimated (simple)  & 2.006  & 1.011    & 1.007    \\
     & 50                        & Estimated (complex) & 3.026  & 1.046    & 1.041    \\
      & 100                       & True                & 6.856  & 0.667    & 0.655    \\
      & 100                       & Estimated (simple)  & 1.431  & 0.667    & 0.654    \\
    & 100                       & Estimated (complex) & 1.442  & 0.670    & 0.662    \\
     & 200                       & True                & 4.853  & 0.470    & 0.465    \\
    & 200                       & Estimated (simple)  & 1.004  & 0.470    & 0.465    \\
      & 200                       & Estimated (complex) & 0.826  & 0.471    & 0.469    \\ \midrule 
$\E \big [\overbar{Y}^{a=0} \big]$       & 50                        & True                & 9.764  & 0.987    & 0.983    \\
    & 50                        & Estimated (simple)  & 1.997  & 0.987    & 0.984    \\
     & 50                        & Estimated (complex) & 3.068  & 1.013    & 1.010    \\
     & 100                       & True                & 6.842  & 0.676    & 0.666    \\
     & 100                       & Estimated (simple)  & 1.389  & 0.676    & 0.666    \\
      & 100                       & Estimated (complex) & 1.436  & 0.681    & 0.674    \\
     & 200                       & True                & 4.859  & 0.473    & 0.465    \\
      & 200                       & Estimated (simple)  & 0.998  & 0.473    & 0.465    \\
      & 200                       & Estimated (complex) & 0.859  & 0.471    & 0.467    \\ \bottomrule
\end{tabular}
\caption*{Results are scaled by $\sqrt{m}$ (i.e., multiplied by $\sqrt{5000} \approx 70.7$). ATE is defined as $\E \big [\overbar{Y}^{a=1} \big] - \E \big [\overbar{Y}^{a=0} \big] $; $n$ = number of clusters in the trial; Values for probabilities = how the treatment and sampling probabilities are obtained for the estimators; True = use the known-by-design sampling probabilities and probabilities of treatment in the trial; Estimated (simple) = estimate the sampling and treatment probabilities conditional on $X_j$ only (the variable used to determine the sampling probabilities); Estimated (complex) = estimate the sampling and treatment probabilities conditional on $X_j$ and the cluster-level averages of $\boldsymbol{W}_{1,j}$ and $\boldsymbol{W}_{2,j}$; IPW = non-augmented inverse probability weighting estimator; AIPW1 = augmented inverse probability weighting estimator, with the outcome model fit only at the cluster-level; AIPW2 = augmented inverse probability weighting estimator, with the outcome model fit at the individual-level.}
\end{table}
\renewcommand{\arraystretch}{1}

\clearpage
\section{Simulation scenario 3}

\renewcommand{\arraystretch}{1.16}
\begin{table}[ht!]
\centering
\caption{Scaled bias of estimators for quantities in the target population.}
\label{table_selected_results_simulation_targetALL}
\small
\begin{tabular}{lcccccc}
\toprule
Estimand & $n$   & Values for probabilities & Trial-only  & IPW     & AIPW1   & AIPW2   \\ \midrule
ATE      & 50                        & True                & 7.388  & -0.031 & 0.021  & 0.022  \\
      & 50                        & Estimated (simple)  & 7.388  & 0.024  & 0.021  & 0.022  \\
      & 50                        & Estimated (complex) & 7.388  & 0.019  & 0.012  & 0.022  \\
      & 100                       & True                & 7.291  & -0.090  & 0.009  & 0.008  \\
      & 100                       & Estimated (simple)  & 7.291  & 0.023  & 0.009  & 0.008  \\
      & 100                       & Estimated (complex) & 7.291  & 0.024  & 0.009  & 0.013  \\
      & 200                       & True                & 7.336  & -0.329 & 0.012  & 0.013  \\
      & 200                       & Estimated (simple)  & 7.336  & -0.013 & 0.012  & 0.012  \\
      & 200                       & Estimated (complex) & 7.336  & -0.034 & 0.007  & 0.010   \\ \midrule
$\E \big [\overbar{Y}^{a=1}  \big]$      & 50                        & True                & 3.667  & -0.113 & 0.012  & 0.012  \\
     & 50                        & Estimated (simple)  & 3.667  & -0.007 & 0.012  & 0.012  \\
     & 50                        & Estimated (complex) & 3.667  & 0.032  & 0.009  & 0.013  \\
     & 100                       & True                & 3.673  & 0.057  & -0.006 & -0.008 \\
     & 100                       & Estimated (simple)  & 3.673  & 0.033  & -0.006 & -0.008 \\
      & 100                       & Estimated (complex) & 3.673  & 0.024  & -0.003 & -0.002 \\
      & 200                       & True                & 3.682  & -0.169 & 0.006  & 0.007  \\
      & 200                       & Estimated (simple)  & 3.682  & 0.001  & 0.006  & 0.007  \\
      & 200                       & Estimated (complex) & 3.682  & -0.020  & 0.004  & 0.006  \\ \midrule 
$\E \big [\overbar{Y}^{a=0}  \big]$        & 50                        & True                & -3.721 & -0.082 & -0.009 & -0.010  \\
      & 50                        & Estimated (simple)  & -3.721 & -0.031 & -0.009 & -0.010  \\
      & 50                        & Estimated (complex) & -3.721 & 0.014  & -0.002 & -0.009 \\
      & 100                       & True                & -3.618 & 0.148  & -0.015 & -0.017 \\
      & 100                       & Estimated (simple)  & -3.618 & 0.010   & -0.015 & -0.016 \\
      & 100                       & Estimated (complex) & -3.618 & 0.000      & -0.012 & -0.015 \\
      & 200                       & True                & -3.654 & 0.160   & -0.006 & -0.006 \\
      & 200                       & Estimated (simple)  & -3.654 & 0.014  & -0.006 & -0.006 \\
      & 200                       & Estimated (complex) & -3.654 & 0.014  & -0.003 & -0.004 \\ \bottomrule 
\end{tabular}
\caption*{Results are scaled by $\sqrt{m}$ (i.e., multiplied by $\sqrt{5000} \approx 70.7$). ATE is defined as $\E \big [\overbar{Y}^{a=1} \big] - \E \big [\overbar{Y}^{a=0} \big] $; $n$ = number of clusters in the trial; Values for probabilities = how the treatment and sampling probabilities are obtained for the estimators (True = use the known-by-design sampling probabilities and probabilities of treatment in the trial; Estimated (simple) = estimate the sampling and treatment probabilities conditional on $X_j$ only (the variable used to determine the sampling probabilities); Estimated (complex) = estimate the sampling and treatment probabilities conditional on $X_j$ and the cluster-level averages of $\boldsymbol{W}_{1,j}$ and $\boldsymbol{W}_{2,j}$); Trial-only = average the individual-level outcomes in each cluster and then take the average of these averages over the clusters participating in the trial; IPW = non-augmented inverse probability weighting estimator; AIPW1 = augmented inverse probability weighting estimator, with the outcome model fit only at the cluster-level; AIPW2 = augmented inverse probability weighting estimator, with the outcome model fit at the individual-level.}
\end{table}
\renewcommand{\arraystretch}{1}

\renewcommand{\arraystretch}{1.16}
\begin{table}[ht!]
\centering
\caption{Scaled standard deviation of estimators for quantities in the target population.}
\label{table_selected_results_simulation_targetALL_SD_appendix_scenario4}
\begin{tabular}{lccccc}
\toprule
Estimand & $n$   & Values for probabilities & IPW & AIPW1 & AIPW2 \\ \midrule
ATE      & 50                        & True                & 13.680  & 1.447    & 1.446    \\
      & 50                        & Estimated (simple)  & 2.976  & 1.447    & 1.447    \\
      & 50                        & Estimated (complex) & 4.180   & 1.492    & 1.490     \\
      & 100                       & True                & 9.536  & 0.975    & 0.973    \\
      & 100                       & Estimated (simple)  & 2.130   & 0.975    & 0.972    \\
      & 100                       & Estimated (complex) & 1.968  & 0.985    & 0.983    \\
      & 200                       & True                & 6.923  & 0.699    & 0.698    \\
      & 200                       & Estimated (simple)  & 1.496  & 0.699    & 0.698    \\
      & 200                       & Estimated (complex) & 1.186  & 0.705    & 0.704    \\ \midrule
$\E \big [\overbar{Y}^{a=1}  \big]$      & 50                        & True                & 9.586  & 1.043    & 1.040     \\
      & 50                        & Estimated (simple)  & 2.128  & 1.043    & 1.040     \\
      & 50                        & Estimated (complex) & 3.106  & 1.080     & 1.076    \\
      & 100                       & True                & 6.883  & 0.679    & 0.675    \\
      & 100                       & Estimated (simple)  & 1.536  & 0.679    & 0.675    \\
      & 100                       & Estimated (complex) & 1.505  & 0.686    & 0.682    \\
      & 200                       & True                & 4.863  & 0.484    & 0.483    \\
      & 200                       & Estimated (simple)  & 1.072  & 0.484    & 0.483    \\
      & 200                       & Estimated (complex) & 0.869  & 0.489    & 0.488    \\ \midrule
$\E \big [\overbar{Y}^{a=0}  \big]$       & 50                        & True                & 9.791  & 1.021    & 1.021    \\
      & 50                        & Estimated (simple)  & 2.128  & 1.021    & 1.022    \\
      & 50                        & Estimated (complex) & 3.110   & 1.043    & 1.044    \\
      & 100                       & True                & 6.835  & 0.681    & 0.680     \\
      & 100                       & Estimated (simple)  & 1.466  & 0.681    & 0.680     \\
      & 100                       & Estimated (complex) & 1.464  & 0.692    & 0.690     \\
      & 200                       & True                & 4.870   & 0.483    & 0.480     \\
      & 200                       & Estimated (simple)  & 1.067  & 0.483    & 0.481    \\
      & 200                       & Estimated (complex) & 0.895  & 0.485    & 0.483    \\ \bottomrule
 \\
        \bottomrule
\end{tabular}
\caption*{Results are scaled by $\sqrt{m}$ (i.e., multiplied by $\sqrt{5000} \approx 70.7$). ATE is defined as $\E \big [\overbar{Y}^{a=1} \big] - \E \big [\overbar{Y}^{a=0} \big] $; $n$ = number of clusters in the trial; Values for probabilities = how the treatment and sampling probabilities are obtained for the estimators; True = use the known-by-design sampling probabilities and probabilities of treatment in the trial; Estimated (simple) = estimate the sampling and treatment probabilities conditional on $X_j$ only (the variable used to determine the sampling probabilities); Estimated (complex) = estimate the sampling and treatment probabilities conditional on $X_j$ and the cluster-level averages of $\boldsymbol{W}_{1,j}$ and $\boldsymbol{W}_{2,j}$; IPW = non-augmented inverse probability weighting estimator; AIPW1 = augmented inverse probability weighting estimator, with the outcome model fit only at the cluster-level; AIPW2 = augmented inverse probability weighting estimator, with the outcome model fit at the individual-level.}
\end{table}
\renewcommand{\arraystretch}{1}

\clearpage
\section{Simulation scenario 4}

\renewcommand{\arraystretch}{1.16}
\begin{table}[ht!]
\centering
\caption{Scaled bias of estimators for quantities in the target population.}
\label{table_selected_results_simulation_targetALL}
\small
\begin{tabular}{lcccccc}
\toprule
Estimand & $n$   & Values for probabilities & Trial-only  & IPW     & AIPW1   & AIPW2   \\ \midrule
ATE      & 50                        & True                & 0.083  & 0.047  & 0.017  & 0.013  \\
      & 50                        & Estimated (simple)  & 0.083  & 0.070   & 0.017  & 0.013  \\
      & 50                        & Estimated (complex) & 0.083  & -0.001 & 0.016  & 0.018  \\
      & 100                       & True                & -0.060  & -0.171 & 0.004  & 0.011  \\
      & 100                       & Estimated (simple)  & -0.060  & -0.046 & 0.004  & 0.011  \\
      & 100                       & Estimated (complex) & -0.060  & -0.020  & 0.009  & 0.013  \\
      & 200                       & True                & -0.040  & -0.319 & -0.005 & -0.005 \\
      & 200                       & Estimated (simple)  & -0.040  & -0.018 & -0.005 & -0.005 \\
      & 200                       & Estimated (complex) & -0.040  & -0.010  & -0.007 & -0.007 \\ \midrule
$\E \big [\overbar{Y}^{a=1}  \big]$      & 50                        & True                & -6.536 & -0.032 & 0.006  & 0.003  \\
      & 50                        & Estimated (simple)  & -6.536 & 0.038  & 0.006  & 0.003  \\
      & 50                        & Estimated (complex) & -6.536 & 0.007  & 0.016  & 0.009  \\
      & 100                       & True                & -6.570  & -0.013 & -0.006 & -0.002 \\
      & 100                       & Estimated (simple)  & -6.570  & -0.036 & -0.006 & -0.002 \\
      & 100                       & Estimated (complex) & -6.570  & -0.018 & 0.000      & 0.000      \\
      & 200                       & True                & -6.596 & -0.160  & -0.005 & -0.006 \\
      & 200                       & Estimated (simple)  & -6.596 & -0.003 & -0.005 & -0.006 \\
      & 200                       & Estimated (complex) & -6.596 & 0.007  & -0.004 & -0.007 \\ \midrule
$\E \big [\overbar{Y}^{a=0}  \big]$        & 50                        & True                & -6.619 & -0.079 & -0.010  & -0.009 \\
      & 50                        & Estimated (simple)  & -6.619 & -0.033 & -0.010  & -0.010  \\
      & 50                        & Estimated (complex) & -6.619 & 0.008  & 0.000      & -0.009 \\
      & 100                       & True                & -6.510  & 0.158  & -0.011 & -0.013 \\
      & 100                       & Estimated (simple)  & -6.510  & 0.010   & -0.011 & -0.013 \\
      & 100                       & Estimated (complex) & -6.510  & 0.002  & -0.009 & -0.013 \\
      & 200                       & True                & -6.555 & 0.158  & 0.000      & -0.001 \\
      & 200                       & Estimated (simple)  & -6.555 & 0.015  & 0.000      & -0.001 \\
      & 200                       & Estimated (complex) & -6.555 & 0.017  & 0.003  & 0.000      \\ \bottomrule
\end{tabular}
\caption*{Results are scaled by $\sqrt{m}$ (i.e., multiplied by $\sqrt{5000} \approx 70.7$). ATE is defined as $\E \big [\overbar{Y}^{a=1} \big] - \E \big [\overbar{Y}^{a=0} \big] $; $n$ = number of clusters in the trial; Values for probabilities = how the treatment and sampling probabilities are obtained for the estimators (True = use the known-by-design sampling probabilities and probabilities of treatment in the trial; Estimated (simple) = estimate the sampling and treatment probabilities conditional on $X_j$ only (the variable used to determine the sampling probabilities); Estimated (complex) = estimate the sampling and treatment probabilities conditional on $X_j$ and the cluster-level averages of $\boldsymbol{W}_{1,j}$ and $\boldsymbol{W}_{2,j}$); Trial-only = average the individual-level outcomes in each cluster and then take the average of these averages over the clusters participating in the trial; IPW = non-augmented inverse probability weighting estimator; AIPW1 = augmented inverse probability weighting estimator, with the outcome model fit only at the cluster-level; AIPW2 = augmented inverse probability weighting estimator, with the outcome model fit at the individual-level.}
\end{table}
\renewcommand{\arraystretch}{1}

\renewcommand{\arraystretch}{1.16}
\begin{table}[ht!]
\centering
\caption{Scaled standard deviation of estimators for quantities in the target population.}
\label{table_selected_results_simulation_targetALL_SD_appendix_scenario4}
\begin{tabular}{lccccc}
\toprule
Estimand & $n$   & Values for probabilities & IPW & AIPW1 & AIPW2 \\ \midrule
ATE      & 50                        & True                & 13.636 & 1.395    & 1.373    \\
      & 50                        & Estimated (simple)  & 2.892  & 1.395    & 1.375    \\
      & 50                        & Estimated (complex) & 4.206  & 1.440     & 1.420     \\
      & 100                       & True                & 9.454  & 0.944    & 0.938    \\
      & 100                       & Estimated (simple)  & 1.952  & 0.944    & 0.937    \\
      & 100                       & Estimated (complex) & 1.985  & 0.950     & 0.943    \\
      & 200                       & True                & 6.922  & 0.640     & 0.632    \\
      & 200                       & Estimated (simple)  & 1.388  & 0.640     & 0.632    \\
      & 200                       & Estimated (complex) & 1.216  & 0.638    & 0.634    \\ \midrule
$\E \big [\overbar{Y}^{a=1}  \big]$       & 50                        & True                & 9.692  & 1.014    & 1.000        \\
      & 50                        & Estimated (simple)  & 2.056  & 1.014    & 1.001    \\
      & 50                        & Estimated (complex) & 3.078  & 1.056    & 1.043    \\
      & 100                       & True                & 6.811  & 0.675    & 0.675    \\
      & 100                       & Estimated (simple)  & 1.373  & 0.675    & 0.675    \\
      & 100                       & Estimated (complex) & 1.473  & 0.683    & 0.682    \\
      & 200                       & True                & 4.909  & 0.466    & 0.460     \\
      & 200                       & Estimated (simple)  & 0.999  & 0.466    & 0.460     \\
      & 200                       & Estimated (complex) & 0.868  & 0.466    & 0.463    \\ \midrule 
$\E \big [\overbar{Y}^{a=0}  \big]$       & 50                        & True                & 9.766  & 0.987    & 0.983    \\
      & 50                        & Estimated (simple)  & 1.997  & 0.987    & 0.984    \\
      & 50                        & Estimated (complex) & 3.069  & 1.013    & 1.010     \\
      & 100                       & True                & 6.842  & 0.676    & 0.666    \\
      & 100                       & Estimated (simple)  & 1.389  & 0.676    & 0.666    \\
      & 100                       & Estimated (complex) & 1.436  & 0.681    & 0.674    \\
      & 200                       & True                & 4.859  & 0.473    & 0.465    \\
      & 200                       & Estimated (simple)  & 0.998  & 0.473    & 0.465    \\
      & 200                       & Estimated (complex) & 0.859  & 0.471    & 0.467    \\ \bottomrule
\end{tabular}
\caption*{Results are scaled by $\sqrt{m}$ (i.e., multiplied by $\sqrt{5000} \approx 70.7$). ATE is defined as $\E \big [\overbar{Y}^{a=1} \big] - \E \big [\overbar{Y}^{a=0} \big] $; $n$ = number of clusters in the trial; Values for probabilities = how the treatment and sampling probabilities are obtained for the estimators; True = use the known-by-design sampling probabilities and probabilities of treatment in the trial; Estimated (simple) = estimate the sampling and treatment probabilities conditional on $X_j$ only (the variable used to determine the sampling probabilities); Estimated (complex) = estimate the sampling and treatment probabilities conditional on $X_j$ and the cluster-level averages of $\boldsymbol{W}_{1,j}$ and $\boldsymbol{W}_{2,j}$; IPW = non-augmented inverse probability weighting estimator; AIPW1 = augmented inverse probability weighting estimator, with the outcome model fit only at the cluster-level; AIPW2 = augmented inverse probability weighting estimator, with the outcome model fit at the individual-level.}
\end{table}
\renewcommand{\arraystretch}{1}

\clearpage
\section{Simulation scenario 5}

\renewcommand{\arraystretch}{1.16}
\begin{table}[ht!]
\centering
\caption{Scaled bias of estimators for quantities in the target population.}
\label{table_selected_results_simulation_targetALL}
\small
\begin{tabular}{lcccccc}
\toprule
Estimand & $n$   & Values for probabilities & Trial-only  & IPW     & AIPW1   & AIPW2   \\ \midrule
ATE      & 50                        & True                & 12.286 & -0.045 & 0.028  & 0.024  \\
     & 50                        & Estimated (simple)  & 12.286 & 0.015  & 0.028  & 0.029  \\
     & 50                        & Estimated (complex) & 12.286 & -0.014 & 0.006  & 0.03   \\
     & 100                       & True                & 12.159 & -0.094 & -0.049 & -0.052 \\
     & 100                       & Estimated (simple)  & 12.159 & 0.025  & -0.049 & -0.053 \\
      & 100                       & Estimated (complex) & 12.159 & -0.009 & -0.062 & -0.054 \\
     & 200                       & True                & 12.225 & -0.345 & 0.002  & 0.004  \\
     & 200                       & Estimated (simple)  & 12.225 & -0.028 & 0.002  & 0.004  \\
     & 200                       & Estimated (complex) & 12.225 & -0.023 & -0.004 & 0.002  \\ \midrule
$\E \big [\overbar{Y}^{a=1}  \big]$       & 50                        & True                & 6.099  & -0.144 & -0.021 & -0.022 \\
    & 50                        & Estimated (simple)  & 6.099  & -0.024 & -0.021 & -0.020  \\
     & 50                        & Estimated (complex) & 6.099  & -0.01  & -0.026 & -0.015 \\
      & 100                       & True                & 6.133  & 0.093  & -0.021 & -0.026 \\
      & 100                       & Estimated (simple)  & 6.133  & 0.070   & -0.021 & -0.026 \\
      & 100                       & Estimated (complex) & 6.133  & 0.031  & -0.026 & -0.022 \\
     & 200                       & True                & 6.138  & -0.175 & -0.008 & -0.009 \\
     & 200                       & Estimated (simple)  & 6.138  & -0.002 & -0.008 & -0.009 \\
      & 200                       & Estimated (complex) & 6.138  & -0.010  & -0.012 & -0.010  \\ \midrule
$\E \big [\overbar{Y}^{a=0}  \big]$         & 50                        & True                & -6.187 & -0.1   & -0.049 & -0.046 \\
     & 50                        & Estimated (simple)  & -6.187 & -0.039 & -0.049 & -0.049 \\
      & 50                        & Estimated (complex) & -6.187 & 0.004  & -0.033 & -0.045 \\
      & 100                       & True                & -6.026 & 0.187  & 0.028  & 0.026  \\
      & 100                       & Estimated (simple)  & -6.026 & 0.045  & 0.028  & 0.027  \\
     & 100                       & Estimated (complex) & -6.026 & 0.04   & 0.036  & 0.031  \\
     & 200                       & True                & -6.086 & 0.170   & -0.010  & -0.013 \\
     & 200                       & Estimated (simple)  & -6.086 & 0.026  & -0.010  & -0.012 \\
     & 200                       & Estimated (complex) & -6.086 & 0.013  & -0.007 & -0.012 \\ \bottomrule
\end{tabular}
\caption*{Results are scaled by $\sqrt{m}$ (i.e., multiplied by $\sqrt{5000} \approx 70.7$). ATE is defined as $\E \big [\overbar{Y}^{a=1} \big] - \E \big [\overbar{Y}^{a=0} \big] $; $n$ = number of clusters in the trial; Values for probabilities = how the treatment and sampling probabilities are obtained for the estimators (True = use the known-by-design sampling probabilities and probabilities of treatment in the trial; Estimated (simple) = estimate the sampling and treatment probabilities conditional on $X_j$ only (the variable used to determine the sampling probabilities); Estimated (complex) = estimate the sampling and treatment probabilities conditional on $X_j$ and the cluster-level averages of $\boldsymbol{W}_{1,j}$ and $\boldsymbol{W}_{2,j}$); Trial-only = average the individual-level outcomes in each cluster and then take the average of these averages over the clusters participating in the trial; IPW = non-augmented inverse probability weighting estimator; AIPW1 = augmented inverse probability weighting estimator, with the outcome model fit only at the cluster-level; AIPW2 = augmented inverse probability weighting estimator, with the outcome model fit at the individual-level.}
\end{table}

\renewcommand{\arraystretch}{1}

\renewcommand{\arraystretch}{1.16}
\begin{table}[ht!]
\centering
\caption{Scaled standard deviation of estimators for quantities in the target population.}
\label{table_selected_results_simulation_targetALL_SD_appendix_scenario5}
\begin{tabular}{lccccc}
\toprule
Estimand & $n$   & Values for probabilities & IPW & AIPW1 & AIPW2 \\ \midrule
ATE      & 50                        & True                & 14.133 & 3.749    & 3.747    \\
     & 50                        & Estimated (simple)  & 4.847  & 3.749    & 3.747    \\
      & 50                        & Estimated (complex) & 4.853  & 3.773    & 3.768    \\
     & 100                       & True                & 9.887  & 2.535    & 2.530    \\
     & 100                       & Estimated (simple)  & 3.412  & 2.535    & 2.529    \\
     & 100                       & Estimated (complex) & 3.087  & 2.553    & 2.546    \\
      & 200                       & True                & 7.152  & 1.752    & 1.743    \\
     & 200                       & Estimated (simple)  & 2.436  & 1.752    & 1.743    \\
     & 200                       & Estimated (complex) & 2.012  & 1.747    & 1.744    \\ \midrule 
$\E \big [\overbar{Y}^{a=1} \big]$        & 50                        & True                & 9.885  & 2.668    & 2.654    \\
      & 50                        & Estimated (simple)  & 3.464  & 2.668    & 2.654    \\
     & 50                        & Estimated (complex) & 3.572  & 2.695    & 2.678    \\
     & 100                       & True                & 7.149  & 1.839    & 1.830    \\
     & 100                       & Estimated (simple)  & 2.458  & 1.839    & 1.830    \\
      & 100                       & Estimated (complex) & 2.246  & 1.852    & 1.843    \\
    & 200                       & True                & 5.014  & 1.247    & 1.240    \\
     & 200                       & Estimated (simple)  & 1.732  & 1.247    & 1.240    \\
     & 200                       & Estimated (complex) & 1.474  & 1.244    & 1.240    \\ \midrule 
$\E \big [\overbar{Y}^{a=0} \big]$        & 50                        & True                & 10.088 & 2.612    & 2.613    \\
     & 50                        & Estimated (simple)  & 3.444  & 2.612    & 2.615    \\
     & 50                        & Estimated (complex) & 3.473  & 2.634    & 2.637    \\
      & 100                       & True                & 7.093  & 1.737    & 1.736    \\
      & 100                       & Estimated (simple)  & 2.364  & 1.737    & 1.734    \\
     & 100                       & Estimated (complex) & 2.165  & 1.750    & 1.746    \\
     & 200                       & True                & 5.036  & 1.247    & 1.241    \\
   & 200                       & Estimated (simple)  & 1.710  & 1.247    & 1.241    \\
      & 200                       & Estimated (complex) & 1.473  & 1.246    & 1.242    \\ \bottomrule
\end{tabular}
\caption*{Results are scaled by $\sqrt{m}$ (i.e., multiplied by $\sqrt{5000} \approx 70.7$). ATE is defined as $\E \big [\overbar{Y}^{a=1} \big] - \E \big [\overbar{Y}^{a=0} \big] $; $n$ = number of clusters in the trial; Values for probabilities = how the treatment and sampling probabilities are obtained for the estimators; True = use the known-by-design sampling probabilities and probabilities of treatment in the trial; Estimated (simple) = estimate the sampling and treatment probabilities conditional on $X_j$ only (the variable used to determine the sampling probabilities); Estimated (complex) = estimate the sampling and treatment probabilities conditional on $X_j$ and the cluster-level averages of $\boldsymbol{W}_{1,j}$ and $\boldsymbol{W}_{2,j}$; IPW = non-augmented inverse probability weighting estimator; AIPW1 = augmented inverse probability weighting estimator, with the outcome model fit only at the cluster-level; AIPW2 = augmented inverse probability weighting estimator, with the outcome model fit at the individual-level.}
\end{table}
\renewcommand{\arraystretch}{1}

\clearpage
\section{Simulation scenario 6}\label{appendix_scenario6}
\renewcommand{\arraystretch}{1.16}
\begin{table}[ht!]
\centering
\caption{Scaled bias of estimators for quantities in the target population.}
\label{table_selected_results_simulation_targetALL}
\small
\begin{tabular}{lcccccc}
\toprule
Estimand & $n$   & Values for probabilities & Trial-only  & IPW     & AIPW1   & AIPW2   \\ \midrule
ATE      & 50                        & True                & 10.616 & -0.110 & -0.253 & -0.323 \\
     & 50                        & Estimated (simple)  & 10.616 & -0.011 & -0.253 & -0.321 \\
     & 50                        & Estimated (complex) & 10.616 & -0.090 & -0.173 & -0.166 \\
      & 100                       & True                & 10.509 & -0.065 & -0.119 & -0.173 \\
     & 100                       & Estimated (simple)  & 10.509 & 0.002  & -0.119 & -0.173 \\
      & 100                       & Estimated (complex) & 10.509 & -0.021 & -0.071 & -0.081 \\
    & 200                       & True                & 10.566 & -0.345 & -0.052 & -0.064 \\
      & 200                       & Estimated (simple)  & 10.566 & -0.029 & -0.052 & -0.064 \\
      & 200                       & Estimated (complex) & 10.566 & -0.071 & -0.031 & -0.021 \\ \midrule 
$\E \big [\overbar{Y}^{a=1} \big]$       & 50                        & True                & 5.258  & -0.185 & -0.143 & -0.181 \\
      & 50                        & Estimated (simple)  & 5.258  & -0.048 & -0.143 & -0.180 \\
      & 50                        & Estimated (complex) & 5.258  & -0.019 & -0.090 & -0.101 \\
     & 100                       & True                & 5.286  & 0.053  & -0.076 & -0.118 \\
      & 100                       & Estimated (simple)  & 5.286  & 0.020  & -0.076 & -0.118 \\
     & 100                       & Estimated (complex) & 5.286  & -0.007 & -0.051 & -0.068 \\
      & 200                       & True                & 5.299  & -0.218 & -0.023 & -0.029 \\
      & 200                       & Estimated (simple)  & 5.299  & -0.003 & -0.023 & -0.029 \\
      & 200                       & Estimated (complex) & 5.299  & -0.039 & -0.013 & -0.008 \\ \midrule 
$\E \big [\overbar{Y}^{a=0} \big]$       & 50                        & True                & -5.358 & -0.076 & 0.110  & 0.142  \\
     & 50                        & Estimated (simple)  & -5.358 & -0.037 & 0.110  & 0.141  \\
      & 50                        & Estimated (complex) & -5.358 & 0.070  & 0.083  & 0.065  \\
      & 100                       & True                & -5.223 & 0.119  & 0.044  & 0.055  \\
      & 100                       & Estimated (simple)  & -5.223 & 0.018  & 0.044  & 0.055  \\
      & 100                       & Estimated (complex) & -5.223 & 0.014  & 0.019  & 0.013  \\
     & 200                       & True                & -5.267 & 0.127  & 0.030  & 0.035  \\
     & 200                       & Estimated (simple)  & -5.267 & 0.026  & 0.030  & 0.035  \\
     & 200                       & Estimated (complex) & -5.267 & 0.031  & 0.018  & 0.013  \\ \bottomrule
\end{tabular}
\caption*{Results are scaled by $\sqrt{m}$ (i.e., multiplied by $\sqrt{5000} \approx 70.7$). ATE is defined as $\E \big [\overbar{Y}^{a=1} \big] - \E \big [\overbar{Y}^{a=0} \big] $; $n$ = number of clusters in the trial; Values for probabilities = how the treatment and sampling probabilities are obtained for the estimators (True = use the known-by-design sampling probabilities and probabilities of treatment in the trial; Estimated (simple) = estimate the sampling and treatment probabilities conditional on $X_j$ only (the variable used to determine the sampling probabilities); Estimated (complex) = estimate the sampling and treatment probabilities conditional on $X_j$ and the cluster-level averages of $\boldsymbol{W}_{1,j}$ and $\boldsymbol{W}_{2,j}$); Trial-only = average the individual-level outcomes in each cluster and then take the average of these averages over the clusters participating in the trial; IPW = non-augmented inverse probability weighting estimator; AIPW1 = augmented inverse probability weighting estimator, with the outcome model fit only at the cluster-level; AIPW2 = augmented inverse probability weighting estimator, with the outcome model fit at the individual-level.}
\end{table}
\renewcommand{\arraystretch}{1}

\renewcommand{\arraystretch}{1.16}
\begin{table}[ht!]
\centering
\caption{Scaled standard deviation of estimators for quantities in the target population.}
\label{table_selected_results_simulation_targetALL_SD_appendix_scenario6}
\begin{tabular}{lccccc}
\toprule
Estimand & $n$   & Values for probabilities & IPW & AIPW1 & AIPW2 \\ \midrule
ATE      & 50                        & True                & 14.411 & 1.855    & 1.744    \\
     & 50                        & Estimated (simple)  & 4.209  & 1.855    & 1.745    \\
     & 50                        & Estimated (complex) & 4.917  & 1.866    & 1.794    \\
     & 100                       & True                & 10.123 & 1.273    & 1.218    \\
     & 100                       & Estimated (simple)  & 2.924  & 1.273    & 1.217    \\
      & 100                       & Estimated (complex) & 2.510  & 1.248    & 1.227    \\
     & 200                       & True                & 7.311  & 0.915    & 0.855    \\
    & 200                       & Estimated (simple)  & 2.121  & 0.915    & 0.855    \\
    & 200                       & Estimated (complex) & 1.543  & 0.879    & 0.859    \\ \midrule 
$\E \big [\overbar{Y}^{a=1} \big]$        & 50                        & True                & 12.374 & 1.335    & 1.244    \\
     & 50                        & Estimated (simple)  & 3.016  & 1.335    & 1.244    \\
      & 50                        & Estimated (complex) & 4.173  & 1.352    & 1.286    \\
     & 100                       & True                & 8.893  & 0.891    & 0.836    \\
     & 100                       & Estimated (simple)  & 2.111  & 0.891    & 0.836    \\
      & 100                       & Estimated (complex) & 2.037  & 0.880    & 0.847    \\
      & 200                       & True                & 6.273  & 0.622    & 0.576    \\
      & 200                       & Estimated (simple)  & 1.506  & 0.622    & 0.576    \\
      & 200                       & Estimated (complex) & 1.186  & 0.597    & 0.578    \\ \midrule 
$\E \big [\overbar{Y}^{a=0} \big]$        & 50                        & True                & 7.391  & 1.310    & 1.225    \\
     & 50                        & Estimated (simple)  & 2.993  & 1.310    & 1.224    \\
      & 50                        & Estimated (complex) & 3.055  & 1.310    & 1.258    \\
     & 100                       & True                & 5.174  & 0.878    & 0.838    \\
      & 100                       & Estimated (simple)  & 2.036  & 0.878    & 0.839    \\
      & 100                       & Estimated (complex) & 1.640  & 0.869    & 0.852    \\
      & 200                       & True                & 3.685  & 0.629    & 0.582    \\
     & 200                       & Estimated (simple)  & 1.485  & 0.629    & 0.582    \\
     & 200                       & Estimated (complex) & 1.095  & 0.607    & 0.587    \\ \bottomrule
\end{tabular}
\caption*{Results are scaled by $\sqrt{m}$ (i.e., multiplied by $\sqrt{5000} \approx 70.7$). ATE is defined as $\E \big [\overbar{Y}^{a=1} \big] - \E \big [\overbar{Y}^{a=0} \big] $; $n$ = number of clusters in the trial; Values for probabilities = how the treatment and sampling probabilities are obtained for the estimators; True = use the known-by-design sampling probabilities and probabilities of treatment in the trial; Estimated (simple) = estimate the sampling and treatment probabilities conditional on $X_j$ only (the variable used to determine the sampling probabilities); Estimated (complex) = estimate the sampling and treatment probabilities conditional on $X_j$ and the cluster-level averages of $\boldsymbol{W}_{1,j}$ and $\boldsymbol{W}_{2,j}$; IPW = non-augmented inverse probability weighting estimator; AIPW1 = augmented inverse probability weighting estimator, with the outcome model fit only at the cluster-level; AIPW2 = augmented inverse probability weighting estimator, with the outcome model fit at the individual-level.}
\end{table}
\renewcommand{\arraystretch}{1}


\end{document}